\def\eps0{\ensuremath{\epsilon _0}}
\def\2pieps0{\ensuremath{2\pi \epsilon _0}}
\def\4pieps0{\ensuremath{4\pi \epsilon _0}}
\def\k{\ensuremath{\displaystyle \frac{1}{4\pi \epsilon _0}}}

\documentclass[twocolumn]{aastex631}

\usepackage{hyperref}
\usepackage{array}
\usepackage{soul}
\newcolumntype{M}[1]{>{\centering\arraybackslash}m{#1}}
\newcolumntype{P}[1]{>{\centering\arraybackslash}p{#1}}

\usepackage{amsmath}
\usepackage{graphicx}

\shorttitle{HST Low Resolution Stellar Library}
\shortauthors{Pal et al.}

\begin{document}

\title{HST Low Resolution Stellar Library}

\author[0000-0002-3077-4037]{Tathagata Pal}
\affiliation{Washington State University \\
1245 Webster Hall \\
Pullman, WA 99163, USA}

\author[0000-0001-8836-3604]{Islam Khan}
\affiliation{Washington State University \\
1245 Webster Hall \\
Pullman, WA 99163, USA}
\affiliation{Haverford College \\
370 Lancaster Ave \\
Haverford, PA 19041, USA}

\author[0000-0003-1388-5525]{Guy Worthey}
\affiliation{Washington State University \\
1245 Webster Hall \\
Pullman, WA 99163, USA}

\author[0000-0003-1388-5525]{Michael D. Gregg}
\affiliation{University of California, Davis \\
517 Physics Building \\
Davis, CA 95616, USA}

\author[0000-0002-7678-2155]{David R. Silva}
\affiliation{The University of Texas at San Antonio \\
College of Sciences, Dean's Office, Suite 3.205\\
One UTSA Circle San Antonio, TX 78249}

\begin{abstract}

In order to provide fundamental stellar spectra that extend into the UV, Hubble Space Telescope's (HST) Space Telescope Imaging Spectrograph (STIS) targeted 556 stars via proposals GO9088, GO9786, GO10222, and GO13776. Exposures through three low resolution gratings provide wavelength coverage from 0.2 $< \lambda <$ 1 $\mu$m at $\lambda/\Delta\lambda\sim$ 1000. The UV grating (G230LB) scatters red light that results in unwanted signal, especially in cool stars. We applied scattered light corrections and flux corrections arising from pointing errors relative to the center of the 0\farcs2 slit based on \cite{2022stis.rept....5W}. We present 513 fully reduced stellar spectra, fluxed, dereddened, and cross-correlated to zero velocity. Because of the broad spectral range, we can simultaneously study H$\alpha$ and Mg II $\lambda$2800, indicators of chromospheric activity. Their behaviors are decoupled. Besides three cool dwarfs and one giant with mild flares in H$\alpha$, only Be stars show strong H$\alpha$ emission. Mg2800 emission, however, strongly anti-correlates with temperature such that warm stars show absorption and stars cooler than 5000 K  universally show chromospheric emission regardless of dwarf/giant status or metallicity. Transformed to Mg2800 flux emerging from the stellar surface, we find a correlation with temperature with approximately symmetric astrophysical scatter. Previous work had indicated a basal level with asymmetric scatter to strong values. The discrepancy is primarily due to our improved treatment of extinction. We confirm statistically significant time variability in Mg2800 strength for one star.

\end{abstract}


\section{Introduction} \label{sec:intro}

Stellar libraries are important tools used in far-flung corners of astronomy and astrophysics. They contain stellar spectra of a number of pre-selected stars in different wavelength regimes (UV, visible, NIR), a variety of spectral resolutions, and with varied attention to flux calibration. Examples include a library of stellar spectra by \cite{1984ApJS...56..257J}, XSHOOTER  \citep{2022A&A...660A..34V}, MILES \citep{2006MNRAS.371..703S}, Indo-US \citep{2004ApJS..152..251V}, IRTF \citep{2013A&A...549A.129C}, ELODIE \citep{1998yCat..41330221S, 2007astro.ph..3658P}, Lick \citep{1994ApJS...94..687W,2014A&A...561A..36W}, and UVES-POP \citep{2003Msngr.114...10B} libraries. Such libraries are often incorporated into stellar population synthesis models. For example, the MILES library \citep{2006MNRAS.371..703S} was used to compute simple stellar population (SSP) SEDs in the optical wavelength range with comprehensive metallicity coverage \citep{2010IAUS..262...65V, 2011A&A...532A..95F}. There are many other examples, such as
\cite{10.1046/j.1365-8711.2003.06897.x,2022A&A...661A..50V,2004A&A...425..881L,2012MNRAS.424..157V,2022MNRAS.511.3198W}. On a star by star basis, libraries can be used to infer stellar parameters like $T_{\rm eff}$, $\mathrm{log} \, g$, and [Fe/H] \citep[e.g.,][]{2011A&A...525A..71W}. Stellar libraries also find application in study of stellar clusters \citep{1996ASPC...98..115A, 2010IAUS..266..304D}. One notable example is the BaSeL 3.1 stellar SED library \citep{1997A&AS..125..229L,1998A&AS..130...65L,2003ASPC..296..238W}. This library is suitable for study of clusters at low metallicities, and has been exploited for the study of globular clusters \citep{bruzual1997matching, weiss1999colour, 1999A&AS..138...19K}, open clusters \citep{1998MNRAS.298..525P, 1999A&A...349..485L}, and blue stragglers \citep{deng1999blue}. When well flux-calibrated, stellar libraries are also very important for characterization and performance evaluation of observational missions like \textit{Gaia} \citep{2002Ap&SS.280...63S, 2002Ap&SS.280...83L}. Several stellar libraries are built into the exposure time calculators for HST and JWST. They even find use in educational products such as the University of Gettysburg's CLEA and VIREO or New Mexico State University's GEAS laboratory software packages to illustrate the trends among stellar spectra.  

Spectral resolution and wavelength coverage vary among the various existing libraries \citep[c.f.~Table 1 of][]{2022A&A...660A..34V}, but none of them extend shortward of 300 nm into the ultraviolet (UV) regime except those of \cite{1983NIUEN..22R....W} and \cite{1990ApJ...364..272F}, who present 172 and 218 stellar spectra, respectively, observed by the International Ultraviolet Explorer (IUE). An important motivation for the present HST-based library is to relieve the relative scarcity of spectral data in the UV.

Study of integrated spectra in the UV allows us access to the hottest stars, which are main sequence turnoff stars with some blue straggler (BS) contribution. For older stellar populations, UV bright populations include blue horizontal branch (BHB) and post-asymptotic giant branch (PAGB) stars \citep{2012AA...538A.143K}. An important goal is to isolate the various main sequences to chart the star formation history (SFH) of the galaxy \citep{10.1093/mnras/stw2231}. The $d \: \! {\rm log \, age}$/$d \: \! \mathrm{log} \, Z=-$3/2 age/metallicity degeneracy \citep{1994ApJS...95..107W} becomes more like $\approx -1/1$ in the UV. In UV, we have an abundance of strong absorption features that help constrain SFH, metallicity, and abundance ratios better \citep{article, article_1, Ponder_1998, 2007ApJ...657.1046C}. Needless to say, if we want to extend the limit on redshift ($z$) for stellar population studies, the UV regime is of utmost importance \citep{Pettini_2000, 2005ApJ...626..680D, 2010ApJ...718L..73V}.

\cite{wu1983prominent} and \cite{1992ApJS...82..197F} gave the first large, systematic spectral library in UV using data from IUE. The library contained spectra of around 218 stars with a spectral resolution of 7\AA. Hubble Space Telescope's (HST's) Space Telescope Imaging Spectrograph (STIS) improves upon IUE in both flux calibration and spectral resolution. Forty O, PAGB, and He-burning stars were observed with STIS to make a hot star spectral library \citep{2018yCat..36150115K}. Made by stitching together spectra from three different gratings, these spectra have wavelength coverage from $\sim$ 2000{\AA} to $\sim$ 10000{\AA} with a resolution of $R\approx \lambda/\Delta \lambda \sim$ 1000. The hot star library was modeled after an earlier effort called the Next Generation Spectral Library (NGSL, \cite{2006hstc.conf..209G}) which has not so far been completely described in the literature. 
The NGSL covers a wide range of stellar parameters, including metallicity \citep{2010AAS...21546302H, 2012AA...538A.143K, 10.1093/mnras/stw2231}. The original proposal was to obtain spectra of close to 600 stars via ``snapshot'' style programs (GO9088, GO9786, GO10222, and GO13776) in which single orbits left stranded between larger programs are exploited for short observations. Spectra of more than half of the stars that were observed (around 374 stars corresponding to proposals GO9088, GO9786, and GO10222) were reduced and made publicly available by \cite{2009ASSP....7..273H}. The main intent of this paper is to provide a reduction of the full library to the community. The spectral quality is improved by applying additional corrections such as scattered light, slit off-center, and dust corrections.

We also investigate the Mg II $\lambda2800$ feature, which is a pair of resonance lines designated by $h$ and $k$. \cite{1981ApJ...244..504B} used high-resolution IUE spectra on F stars to chart four origins for Mg2800 profile morphology: the main, broad stellar absorption feature, a narrower chromospheric emission core, a rare, even narrower self-absorption, and interstellar absorption. \cite{1990ApJ...364..272F} also noted that, when in emission, it probably indicates a chromospheric origin. \citet{1978ApJ...220..619L} argue that most of the Ca II emission ($\lambda\lambda$3933, 3968) arises in the lower chromosphere, Mg II in the middle chromosphere, and Ly$\alpha$ in the upper chromosphere, and that, together, these resonance features provide the bulk of the radiative cooling that occurs in the layers exterior to the photosphere.

The dynamo action brought about by differential stellar rotation is one of the most commonly accepted mechanisms for magnetic field generations in main sequence stars \citep{1987ARA&A..25..271H, 2012A&A...543A.146F, 10.1093/mnras/sty770}. Chromospheric activity is generally associated with strong magnetic fields \citep{1982AcA....32..263M, 2022MNRAS.514.4300B}. Since stellar rotation is expected to slow down over the lifetime of a star, activity can be presumed to decrease \citep{1988ApJ...334..436B}. This leads to the possibility that chromospheric activity indicators (CaII, MgII, or H$\alpha$) may provide relatively precise chronometric information, at least in predefined spectral type bands \citep{1988ApJ...334..436B}. Although, in general ages would be poorly constrained \citep{2013A&A...551L...8P}. In addition, acoustic shocks without a magnetohydrodynamic component may also contribute to the chromospheric activity \citep{1998ApJ...494..700B,10.1111/j.1365-2966.2011.18421.x,10.1093/mnras/stu1706}.

Connections between Mg2800 strengths and the astrophysics of stellar properties are still tenuous, but could eventually lead to realistic chromosphere models as a function of stellar type and magnetic field strength. In the meantime, we have various empirical clues. \cite{1997A&A...326.1143H} finds that, at spectral type M1, metal-deficient stars are also activity-deficient. \cite{1991AJ....101..655S} compares Mg2800 with the Ca II S index \citep{Vaughan_1980} which measures the width of the emission rather than its strength. They also find that the available sample of 20 FGK stars can be separated into ``high activity'' and ``low activity'' groups at an approximately 4:16 ratio, but that Mg2800 displays a large range of values even amongst the low activity group \citep{10.1111/j.1365-2966.2011.18421.x,10.1093/mnras/stu1706}. Interstellar absorption usually dominates in OB stars \citep{2018A&A...615A.115K}. Due to its wide coverage of parameter spaces, the present library can confirm or extend these trends.

This paper is organized as follows. We describe the observations and sample in $\S$\ref{sec:obs}. The data reduction process is detailed in $\S$\ref{sec:qc}, and additional corrections that affect the continuum shape in $\S$\ref{sec:corrections}. The format of the data catalog is described in $\S$\ref{sec:data}. In $\S$\ref{sec:mg}, we investigate the Mg II 2800 feature and chart the systematics of chromospheric activity across the H-R diagram. We conclude in $\S$\ref{sec:conclusion} with a summary of the results and a discussion of their implications.

\section{Observations and Sample} \label{sec:obs}

\begin{figure}[!ht]
\centering
\includegraphics[width=8.6cm]{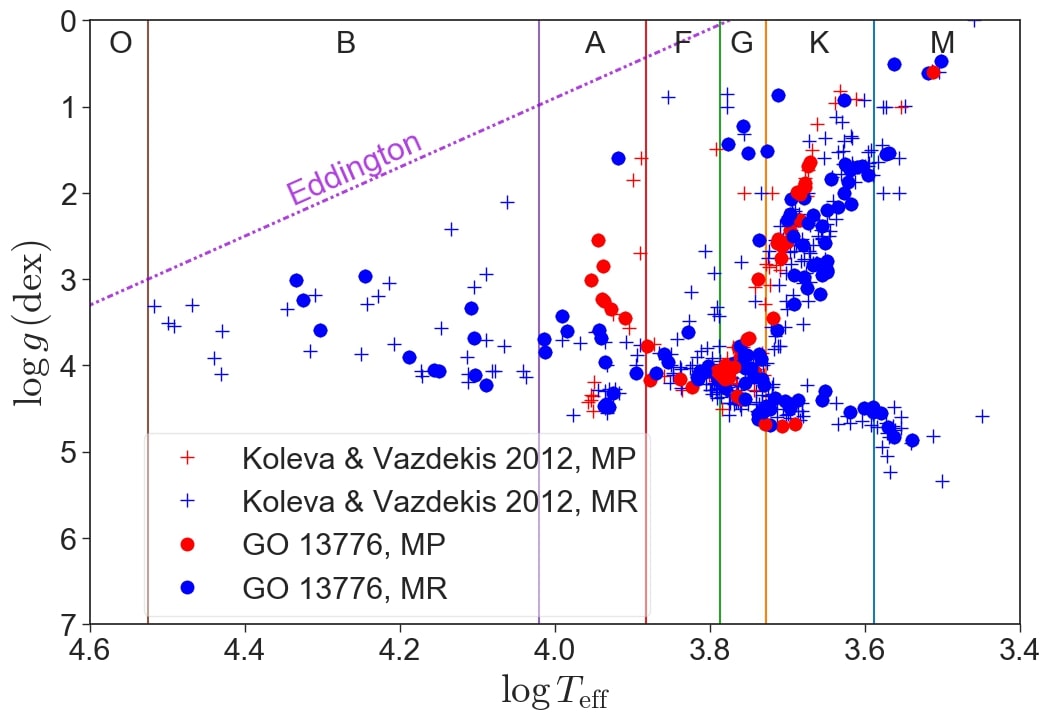}
\caption{NGSL stars are plotted in $\mathrm{log} \, T_{\rm eff}$, $\mathrm{log} \, g$ space. The previously published stars from proposals GO9088, GO9786, and GO10222 \citep[pluses]{2012AA...538A.143K} and the GO13776 stars (circles) are split by metallicity, metal poor (MP): [Fe/H] $\leq$ $-$1 (red) or metal rich (MR): [Fe/H] $>$ $-$1 (blue). An approximate Eddington stability line and spectral type boundaries are included in the plot as visual guides.}
\label{fig:NGSL}
\end{figure}

\begin{figure}[!ht]
\centering
\includegraphics[width=8.6cm]{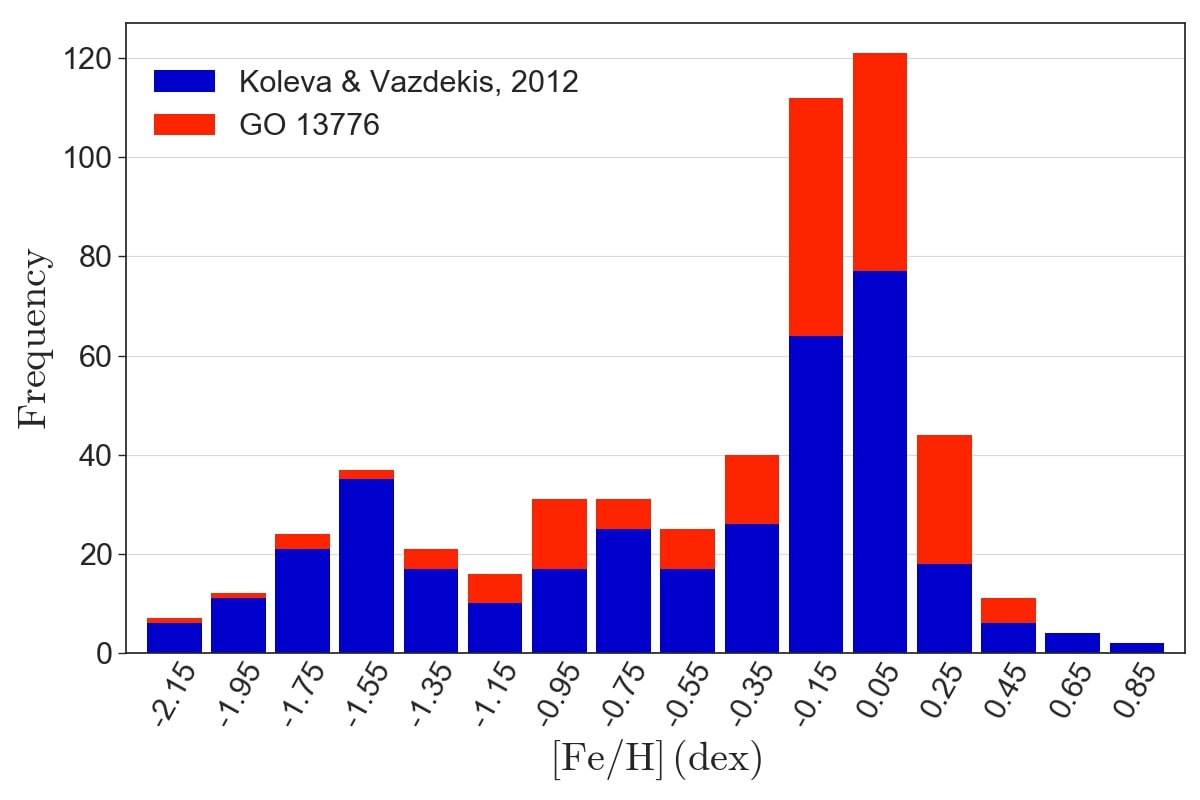}
\caption{The [Fe/H] distribution of all reduced targets in a stacked histogram. Blue corresponds to 345 targets from \cite{2012AA...538A.143K} and red corresponds to 169 targets from HST proposal GO13776. } 
\label{fig:met}
\end{figure}

The stars in the library were selected to cover $T\rm_{eff}$-$L$-$Z$ space insofar as the Galaxy could provide them. For example, given that the metal-poor components of the Milky Way are also ancient in age, no luminous, low-metallicity stars exist. The parameter coverage is shown in Fig.~\ref{fig:NGSL} in $\mathrm{log} \, g$-$\mathrm{log} \, T_{\rm eff}$ space with metallicity indicated by symbol type. Fig.~\ref{fig:met} on the other hand shows the distribution of all the stars in different metallicity bins. The impact of including stars from GO13776 significantly improves coverage of  $T_{\rm eff}$-$L$-$Z$ space, as shown in Figs.~\ref{fig:NGSL} and \ref{fig:met}. Several A-type field horizontal branch stars were observed to attempt to fill in the warm-and-metal-poor gap. Desirable faint stars, such as individual Small Magellanic Cloud stars, could not be observed due to the one-orbit limit on exposure time. The target list is hand-selected, and should not be used for any statistical inferences. In addition, HST's SNAP mode selects from a larger input list according to schedulability, leading to further randomization.

During the orbit in which they were targeted, the NGSL stars were observed by cycling through three different gratings. G230LB sees in UV (central wavelength of 2375\AA), G430L sees in blue (central wavelength of 4300\AA) and G750L sees in red (central wavelength of 7751\AA). The three gratings overlap at 2990\AA-3060{\AA} and 5500\AA-5650\AA \citep{2006hstc.conf..209G}. The CCD detector was employed for these observations. UV exposure times were longer than exposures in the blue or red. A $0 \farcs 2$ slit, equivalent to  $\pm 2$ pixels \citep{2012stii.book.....H,2022stii.book...21P} was used for all the observations and a fringe flat was taken for the G750L grating at the end of each sequence of exposures. 

\begin{figure}[t!]
\centering
\includegraphics[width=8.6cm]{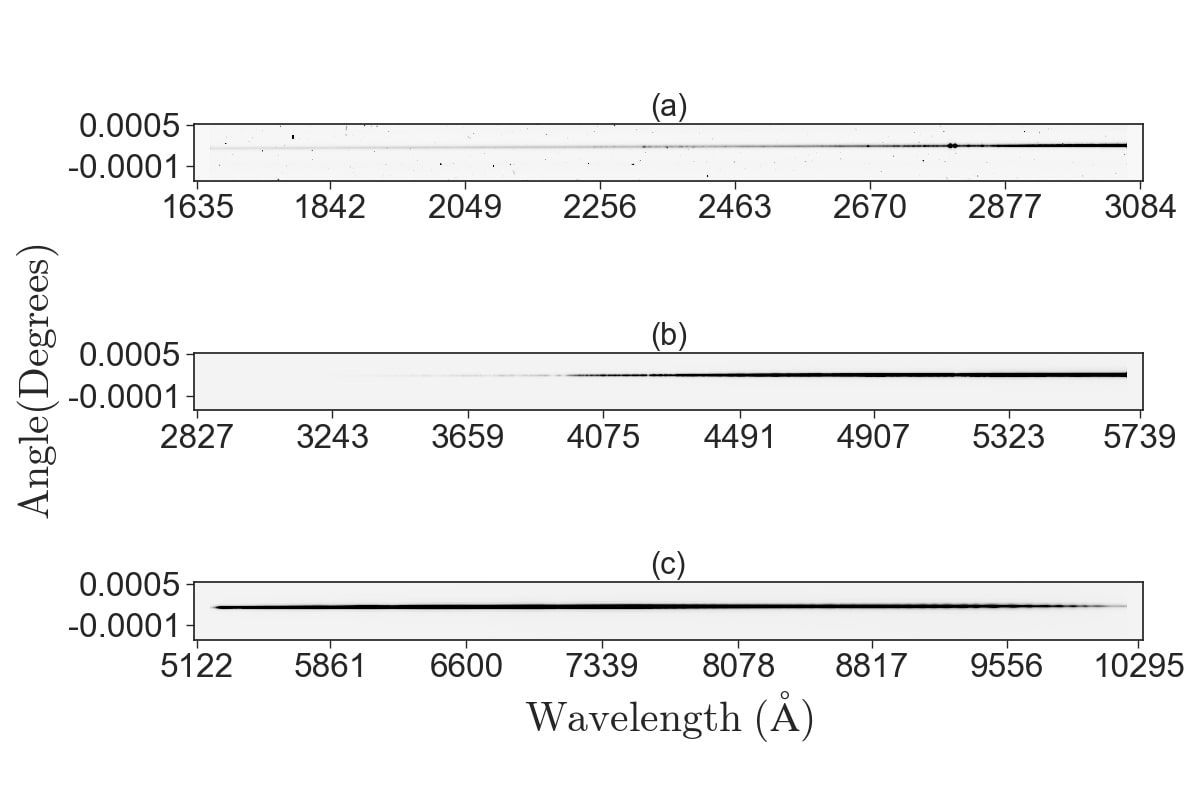}
\caption{CCD images of HD102212 using (a) G230LB, (b) G430L, and (c) G750L. Due to longer exposure time in the UV, the topmost panel shows the presence of cosmic ray events whereas the bottom two do not have any significant ion contamination. It is worth noting for this cool star that what appears to be a stellar trace in the UV shortward of 2500\AA\ is actually light scattered from the visible portion of the spectrum into the UV by grating G230LB. }
\label{fig:uncalib}
\end{figure}

In addition to the usual observational defects (cosmic ray hits, charge transfer efficiency effects, bad pixels, and photon noise) these data suffer from two additional sources of error that affect fluxing.
Firstly, the G230LB grating scatters red light into the UV, creating a spurious signal that must be corrected \citep{2010AAS...21546317L,2022stis.rept....5W}. Secondly, scatter in telescope pointing plus a narrow slit led to situations in which the jaws of the slit sliced off portions of the PSF. Because STIS is an off-axis instrument, the PSF is not symmetrical, so the resultant attenuation is wavelength-dependent. Fortunately, both of these effects can be modeled, and we give details in $\S4$.

Of minor note, STIS spectral flux calibrations have improved since the previous version of the NGSL library was placed at MAST.

\section{Reduction and Quality Control} \label{sec:qc}

All 556 targets from proposals GO9088, GO9786, GO10222, and GO13776 were reduced from raw observation files. Out of these 556 targets, 514 have been reduced completely and additional corrections have been applied. The remaining 42 targets have not been reduced either because of faulty fringe-flat files or because of the absence of one of the observations in UV, blue, or red. The raw files for all the observations (which include observations in UV, blue, and red as well as CCD flats) were downloaded from the Space Telescope Science Institute (STScI) archive. The reduction process is carried on using the \textit{stistools} \textit{Python3} package developed by STScI. Out of these 514 targets, there is a duplication for one of the targets (GJ~614 and HD~145675) leaving us with 513 unique targets.

The reduction procedure consisted of several steps starting from cosmic rays correction to combining disparate spectral windows into one continuous spectrum for each star.

\subsection{Cosmic Ray Correction}\label{subsec:crj}

Cosmic ray corrections are more crucial for observations using G230LB grating that was used for longer-duration UV observations. This is illustrated in Fig.~\ref{fig:uncalib} where cosmic rays are common in the G230LB exposure. Accordingly, all multiple UV observations were run through the \textit{ocrreject} function of \textit{stistools}. This function combined two sets of science observations in UV into a single file. In order to run \textit{ocrreject}, we needed to have at least two observations at each pointing. Unfortunately, the UV raw files from proposals GO10222 and GO13776 did not have multiple UV exposures. For these, bad pixels were removed manually from the spectra.

\subsection{Defringing in the Red}\label{subsec:defringe}

Fringes are interference patterns caused by photons with wavelengths that are integral multiples of the width of the CCD layer. In STIS, fringe patterns are prominent redward of $\sim$7000\AA\ and reach peak-to-peak amplitude of 25\% at 9800\AA\ \citep{Kimble_1998, 2003PASP..115..218M}. The G750L grating produces unwanted fringe patterns. Once per orbit, a fringe flat was obtained using the tungsten lamp on board HST. 

The defringing process was carried out using the \textit{defringe} tool of \textit{stistools} (for details, see \href{https://stistools.readthedocs.io/en/latest/defringe_examples.html}{https://stistools.readthedocs.io/en/latest/}). The following three methods were used in sequence for all the NGSL observations.
\begin{enumerate}
    \item \textit{normspflat}: this method normalizes the fringe-flat that is associated with each observation
    \item \textit{mkfringeflat}: this method cross correlates the normalized fringe-flat with that of the observed spectrum to match the fringes between the two. It minimizes the RMS within a given range of shift and scale values to find the best shift and scale
    \item \textit{defringe}: this method actually defringes the observed spectrum by removing the fringing pattern from the observed spectrum using the shifted and scaled fringe-flat
\end{enumerate} 

\begin{figure}[!ht]
\centering
\includegraphics[width=8.6cm]{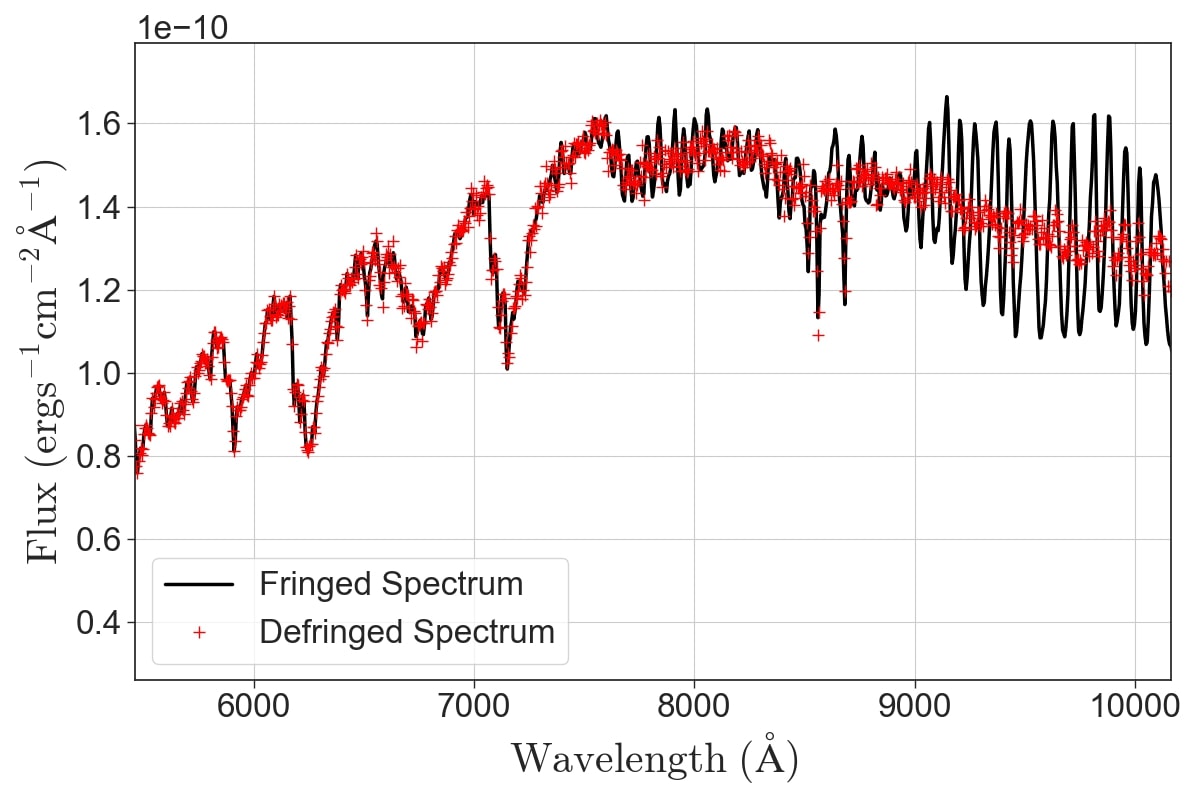}
\caption{Extracted, fluxed CCD/G750L spectrum of HD102212. The spectrum before defringing (black) is compared to the same spectrum after (red). }
\label{fig:fringe}
\end{figure}

Fig.~\ref{fig:fringe} shows the red spectrum of HD102212 before and after defringing.

While defringing the red spectra it was observed that no proper fringe-flat is available for 27 targets. We dropped these stars from further analysis and thus reduced the total number of targets from 556 to 529. While trying to defringe red spectra from GO13776. although some of the targets from GO13776 have 2 or 3 red spectra, only one of them defringed properly. Investigation yielded an observing irregularity. For run GO13776, the G750L (red third of the spectrum) target exposures were preceded by a fringe flat through the $0.3\times0.09$ notch aperture, which is placed near row 512 of the chip (the UV and blue spectra were taken at the E1 pseudoaperture around row 900 of the CCD). The telescope was slewed to place the target star at row 512 of the chip rather than 900, and one exposure taken through the nominal $52\times0.2$ aperture. Due to an oversight, positional dithering occurred. The telescope was slewed $0\farcs5$, and an exposure was taken through the $52\times0.2$ aperture followed by an exposure through the $52\times0.5$ aperture. This last exposure eliminates edge effects and provides the best fluxing, but it cannot be fringe-corrected using the data collected on-orbit. Therefore, only one red spectrum (for each target) was used for run GO13776. For three targets from  GO13776 (HD~65589, HD~84035, and HD~185264), none of the red observations could be satisfactorily defringed. These stars were also dropped from further analysis which reduced the total number of stars from 529 to 526.

Also unique to proposal GO13776, the last pair of red observations were often a pair obtained through 0\farcs2 and 0\farcs5 apertures. Although these could not be defringed due to the shift along the aperture center line, they could be used to create a relative flux correction, should the star have been placed off the central line of the entrance aperture. A smoothed division of these two spectra was applied to the first, defringed observation in all cases where the complete set of observations exists.

\subsection{1-D Extraction}\label{subsec:1d_extraction}

The final step in the reduction process was to extract the 1-D spectrum for each target and each observation in UV, blue, and red using the \textit{x1d} function of \textit{stistools}. This resulted in a separate file for each UV, blue, or red observation for each target. Twelve of the remaining 526 targets did not have one of either UV or blue or red observations. These stars were dropped. This reduced the total number of available stars to 514. 278 targets have 2 observations each of UV, blue, and red. 189 targets have 2 observations each of UV and blue, and 1 of red. The remaining 47 targets have varied numbers of observations for UV, blue and red (at least 1 of each). 

\subsection{Bad Pixel Handling}\label{subsec:markpix}

\begin{figure}[!ht]
\centering
\includegraphics[width=8.6cm]{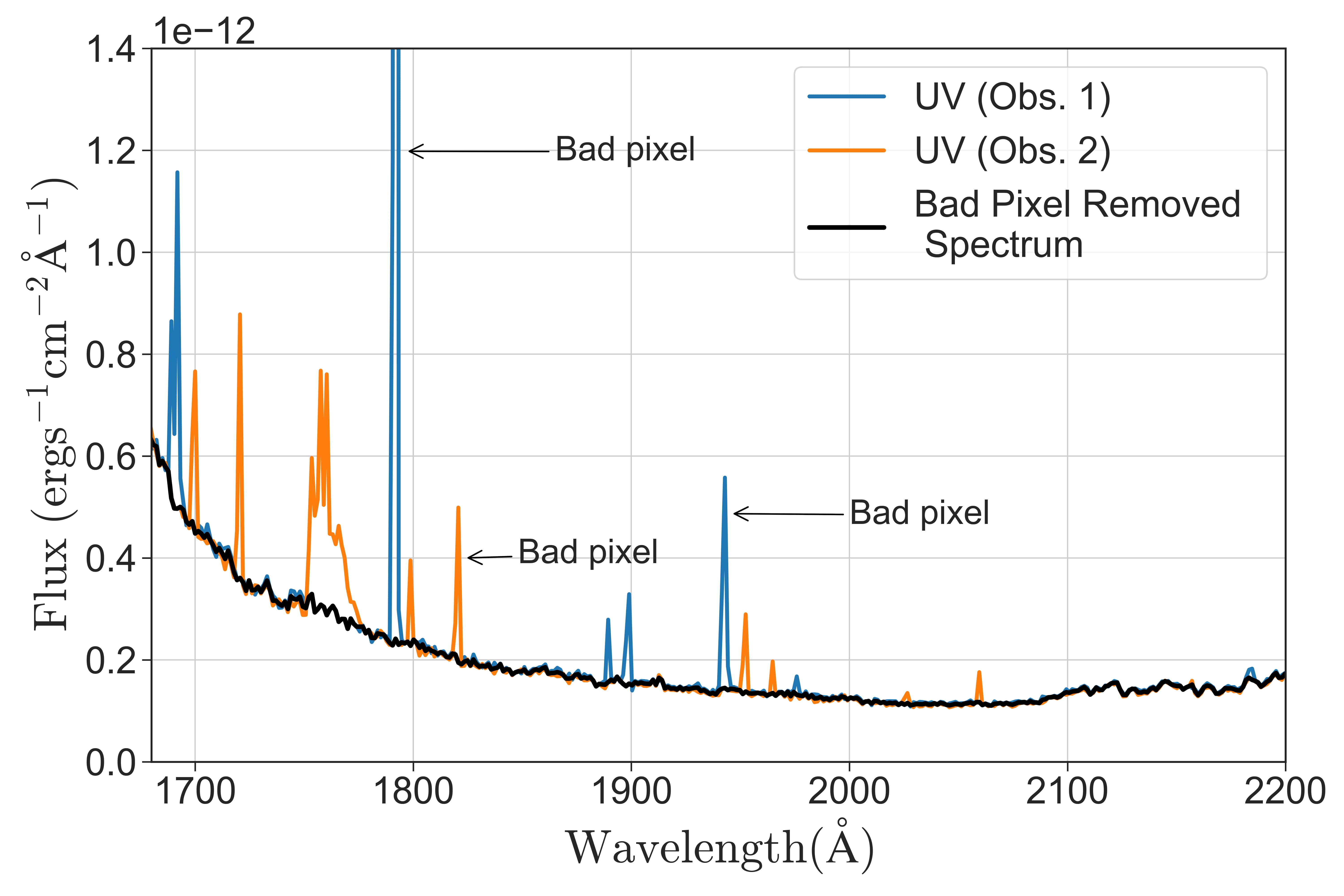}
\caption{Individual spectra for NGSL star HD190360 in the UV (blue and orange) illustrate the presence of bad pixels. After marking, the bad pixels were removed by the algorithm described in $\S$\ref{subsec:composite}. The cleaned spectrum is also plotted (black). We elevated the errors for the corrected portions of the spectrum. }
\label{fig:badpixel}
\end{figure}

As mentioned in Sec.~\ref{subsec:crj}, a cosmic ray rejection algorithm was not applied to blue and red observations. Even after applying cosmic ray rejection to the UV observations, the UV spectra had leftover wild pixels of unusually high and non-astrophysical flux (or counts). In order to mitigate this problem, each observation for each target was checked manually for bad pixels and those pixel locations were flagged. This step generated a single text file for each target containing information on the number of bad pixels for each observation and values for those pixels. Fig.~\ref{fig:badpixel} shows an example of presence of bad pixels in the spectrum.

We tried our best to remove as many bad pixels as possible from each spectrum, but there are some stars for which many bad pixels could not be removed cleanly. 

\subsection{Special Flux Scalings}

After fluxing, some spectra appeared to have been scaled in comparison with their neighbors. For example, suppose a star has been exposed twice in the UV, twice in the blue, and twice in the red. Now and then, one of those six exposures appears slightly too strong or too weak compared with either the spectral overlap region or with its supposedly identical sister spectrum. A scaling was applied to these deviant cases, as listed in Table \ref{tab:scale}. The dataset labels relate closely to the ones assigned by STScI, but we prepended a short string to indicate if the spectrum was UV (uv\_), blue (b\_), or red (r\_).

\begin{deluxetable}{lrrr}
\tabletypesize{\scriptsize}
\tablecaption{Special Scalings}

\tablehead{\colhead{Target} & \colhead{Deviant} & \colhead{Clean} & \colhead{Scale}  \\ 
\colhead{} & \colhead{Dataset} & \colhead{Dataset} & \colhead{Factor}  } 

\startdata
HD 224801 & b\_o93a6qk2q\_flt & b\_o93a6qk3q\_flt  & 1.0506  \\
BD+17 4708 & r\_o6h03vawq\_drj & r\_o6h03vavq\_drj & 1.0669   \\
HD 3712 & r\_o6h04kf0q\_drj & r\_o6h04kezq\_drj    & 1.2163   \\
HD 137759 & r\_o6h04bm3q\_drj & r\_o6h04bm2q\_drj  & 1.1468   \\
HD 124547 & r\_o6h038xkq\_drj & r\_o6h038xjq\_drj  & 1.0556   \\
HD 172506 & r\_o6h06jp4q\_drj & r\_o6h06jp3q\_drj  & 1.0639   \\
HD 4128 & r\_o6h04ynyq\_drj & r\_o6h04ynxq\_drj    & 1.0718   \\
HD 146233 & r\_o6h05wb0q\_drj & r\_o6h05wazq\_drj  & 1.0512   \\
HD 81797 & b\_o6h03rocq\_flt & b\_o6h03robq\_flt   & 1.1058   \\
HD 30614 & uv\_o8ru4c020\_crj & uv\_o8ru4c010\_crj & 0.9720   \\
HR 753 & b\_o6h03ntyq\_flt & b\_o6h03ntzq\_flt     & 1.1994   \\
HD 136442 & b\_ocr7nwr6q\_flt & b\_ocr7nwrcq\_flt  & 0.9319   \\
HD 58343 & uv\_o8ru4s010\_crj & uv\_o8ru4s020\_crj & 0.9668   \\
HD 217014 & b\_ocr7pxp7q\_flt & b\_ocr7pxp6q\_flt  & 0.9346   \\
HD 144608 & r\_ocr7feacq\_drj & b\_ocr7fea7q\_flt  & 0.9048   \\
HD 183324 & b\_o8ruclpqq\_flt & b\_o8ruclprq\_flt  & 1.0501   \\
BD+37 1458 & b\_o6h04ti6q\_flt & b\_o6h04ti7q\_flt & 1.0302   \\
HD 52089 & uv\_o8ru46020\_crj & uv\_o8ru46010\_crj & 0.9725   \\
BD+29 366 & r\_ocr7aif7q\_drj & b\_ocr7aif6q\_flt & 0.947   \\
BD+25 1981 & r\_ocr7agwlq\_drj & b\_ocr7agwkq\_flt & 0.9249 \\
HD 9826 & r\_ocr7kchgq\_drj & b\_ocr7kcheq\_flt & 0.9354 \\
HD 19994 & r\_ocr7klq6q\_drj & b\_ocr7klq4q\_flt & 0.852 \\
HD 21019 & r\_ocr7koizq\_drj & b\_ocr7koiyq\_flt & 0.7542 \\
HD 21770 & r\_ocr7kpsuq\_drj & b\_ocr7kpssq\_flt & 0.8409 \\
HD 25457 & r\_ocr7ksc9q\_drj & b\_ocr7ksc8q\_flt & 0.7998 \\
HD 31128 & r\_ocr7hxziq\_drj & b\_ocr7hxzgq\_flt & 0.9685 \\
HD 34411 & r\_ocr7kxklq\_drj & b\_ocr7kxkkq\_flt & 0.9246 \\
HD 44420 & r\_ocr7lgwsq\_drj & b\_ocr7lgwrq\_flt & 0.9174 \\
HD 48737 & r\_ocr7liuiq\_drj & b\_ocr7liuhq\_flt & 0.9549 \\
HD 52265 & r\_ocr7lln2q\_drj & b\_ocr7lln1q\_flt & 0.9594 \\
HD 57118 & r\_ocr7cqqaq\_drj & b\_ocr7cqq9q\_flt & 0.9343 \\
HD 67523 & r\_ocr7ien9q\_drj & b\_ocr7ien8q\_flt & 0.8912 \\
HD 71369 & r\_ocr7lrsqq\_drj & b\_ocr7lrspq\_flt & 0.9432 \\
HD 82328 & r\_ocr7lyh7q\_drj & b\_ocr7lyh6q\_flt & 0.9042 \\
HD 121370 & r\_ocr7erjeq\_drj & b\_ocr7erjdq\_flt & 0.9313 \\
HD 134169 & r\_ocr7ezp9q\_drj & b\_ocr7ezp8q\_flt & 0.9649 \\
HD 160365 & r\_ocr7odh7q\_drj & b\_ocr7odh6q\_flt & 0.9293 \\
HD 161797 & r\_ocr7oeobq\_drj & b\_ocr7oeoaq\_flt & 0.9371 \\
HD 188510 & r\_ocr7gff0q\_drj & b\_ocr7gfexq\_flt & 0.9354 \\
HD 190390 & r\_ocr7ghheq\_drj & b\_ocr7ghhdq\_flt & 0.939 \\
HD 192718 & r\_ocr7gkaeq\_drj & b\_ocr7gkadq\_flt & 0.9066 \\
HD 217014 & r\_ocr7pxp8q\_drj & b\_ocr7pxp7q\_flt & 0.8636
\enddata

\tablecomments{Additionally, for BD+17 2844 we averaged the red spectra, and for HD 183324 we scaled up both the UV spectra by a factor of 1.093 to match the blue spectra}
\label{tab:scale}
\end{deluxetable}

In addition to sporadic scaling issues, observations for HD 1638 may have missed the target altogether, as all spectral segments contain mostly noise.

\subsection{Relative Velocities and Template Matching}

The NGSL stars were chosen to encompass a broad interval of [Fe/H], $\mathrm{log} \, g$, and $T_{\rm eff}$ \citep{article_ngsl}. Galactic halo stars are mostly metal poor but can possess high relative velocity with respect to the local rest frame \citep{2018ApJ...863...87D}. Thus, some of the stars in NGSL have relative velocities $>250$ km s$^{-1}$. This fact called for a relative velocity correction before bringing all the spectra to rest frame. To be consistent, we applied the relative velocity correction to all 514 stars even when the effects would be negligible. The nonrelativistic formula was used to correct for the relative velocity: 
\begin{equation}\label{rel_vel}
    d\lambda=\frac{v}{c}\times\lambda \, ,
\end{equation} 

where $d\lambda$ is correction to the wavelength $\lambda$, $v$ is the relative velocity of the star in km s$^{-1}$ and $c$ is the speed of light in km s$^{-1}$. $d\lambda$ was added or subtracted from corresponding $\lambda$ values depending on the sign of $v$. The values of $v$ were obtained from the SIMBAD astronomical database \citep{2000A&AS..143....9W}. 

After correcting for the relative velocities, residual shifts to rest frame (vacuum wavelengths) were estimated by comparing with template spectra. The choice of template spectrum was made based on the effective temperature of the particular star. The high resolution templates were rebinned to match the observed wavelength points, then cross-correlated. The following templates were adopted.
\begin{enumerate}
    \item Synthetic spectra were used for cool stars ($T_{\rm eff}<$ 5000 K) and warm stars (5000 K $<$ $T_{\rm eff} <$ 8000 K). The synthetic spectra were generated using \cite{1994ApJS...95..107W} model of stellar population.
    \item The observed spectrum of $\alpha$ Lyrae was used for hot stars ($T_{\rm eff}>$ 8000 K)
\end{enumerate}

\begin{figure}[!ht]
\centering
\includegraphics[width=8.6cm]{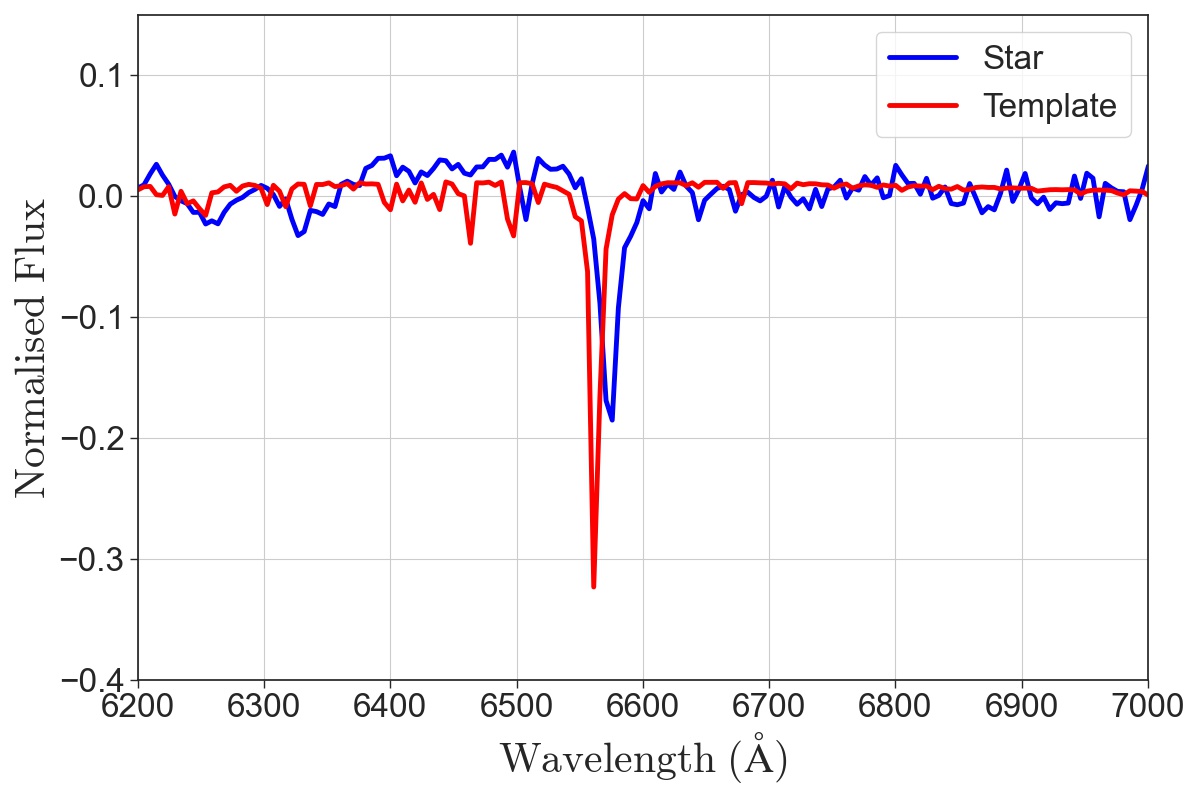}
\caption{A part of spectrum for HD115383 (blue) showing shift of the spectrum with respect to the template (red)}
\label{fig:shift}
\end{figure}

\begin{figure}[!ht]
\centering
\includegraphics[width=8.6cm]{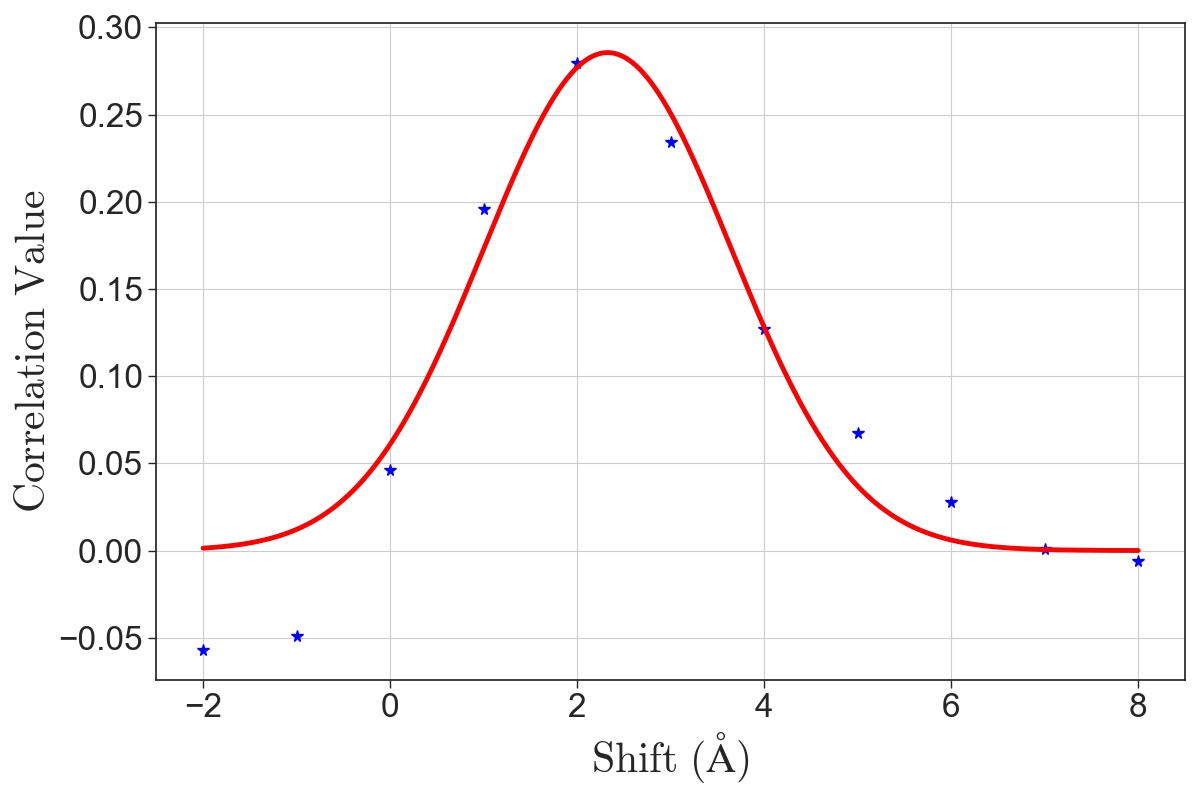}
\caption{Typical cross correlation value as a function of pixel shift in \AA, in this case for the red spectrum of G0 V star HD~115383. }
\label{fig:corr}
\end{figure}

The cross correlation function (in \AA) was fitted with a single peak Gaussian function. Fig.~\ref{fig:shift} shows a part of the spectrum for HD~102212 and illustrates the amount of shift present in the observed spectrum with respect to the template. Correlation value as a function of shift is shown in Fig.~\ref{fig:corr} (for the same star HD~102212). The same template was used for all the observations of a particular target. To speed convergence, we added initial shifts of 3\AA, 9\AA\, and 14\AA\ to UV, blue and red observations, respectively. This ``pre-shift'' evidently arises because wavelength calibrations were not performed on-orbit for NGSL, and so a default wavelength solution was assigned. 

\subsection{The Composite Spectrum}\label{subsec:composite}

To assemble a single contiguous spectrum, we combined bad pixel information and shift information from template matching to splice all the observations for a particular target into one final spectrum. The shift obtained for each observation was added algebraically to the wavelength values. While applying the bad pixel information, we devised a method for suppressing the bad pixels. We first divided the range of each observation into 50 overlapping boxes of 40 pixels each. For each box, we found out the average flux weighted by the variance ($f_{box}$) using the following formula--
\begin{equation}\label{flux_weight}
    f_{box}=\frac{f_1v_1+f_2v_2+...+f_{40}v_{40}}{v_1+v_1+...+v_{40}} \, ,
\end{equation} 

where $f_n$ is the flux at $n^{th}$ wavelength value for a particular box and $v_n$ is the corresponding variance (defined by, $v_n=1/e_{n}^{2}$ where $e_n$ is corresponding error in flux for that particular wavelength value). These flux values were then linearly fitted over the range of observation. Now, the flux at the previously identified bad pixels was set to a flux value according to this linearly extrapolated relation. It is to be noted that the error values at the bad pixels were inflated by a factor of 1000 before calculating $f_{box}$. This was done to make sure that the erroneous pixels do not contribute much to the weighted average (as bad pixels generally have very high flux values).

Once the flux values at the bad pixels were set according to the above mentioned algorithm, we then calculated the weighted average flux value for all the observations of a particular type (for eg., UV, blue or red) at a particular wavelength value. For eg., if there are 2 UV observations for a particular target, then the average UV flux at $n^{th}$ wavelength value ($f^{UV}_{n}$) is given by--
\begin{equation}
    f^{UV}_{n}=\frac{f^{1}_{n}v^{1}_{n}+f^{2}_{n}v^{2}_{n}}{v^{1}_{n}+v^{2}_{n}} \, , 
\end{equation} 

where $f^{1}_{n}$ and $f^{2}_{n}$ are UV fluxes at $n^{th}$ wavelength value for $1^{st}$ and $2^{nd}$ observations respectively and $v^{1}_{n}$ \& $v^{2}_{n}$ are corresponding variances as defined before. This formula can easily be generalized for more than or less than 2 observations. Once this operation was performed for all the observations of a target, we then combined all the observations to make a single spectrum for a target treating $\lambda<$3057\AA\ as UV observation, 3057\AA$<\lambda<$5679\AA\ as blue observation and $\lambda>$5679\AA\ as red observation. This algorithm does not apply without any caveat as sometimes the flux values at bad pixels were negative. Users are advised to be careful of such artifacts in the spectrum by considering the uncertainty we assign.

\section{Continuum Corrections} \label{sec:corrections}

The G230LB grating scatters some red light onto the portions of the CCD where UV is expected \citep{2022stis.rept....5W}. This is a problem mainly for cool stars ($T_{\rm eff} \leq 5000$ K) where we do not expect significant UV flux. This section summarizes the results from \cite{2022stis.rept....5W} on scattered light as well as slit off-center corrections. We also applied these corrections to the 514 NGSL stars that we have reduced.

\subsection{Scattered Light Correction}\label{subsec:scattered_light}
\begin{figure}
\centering
\includegraphics[width=8.6cm]{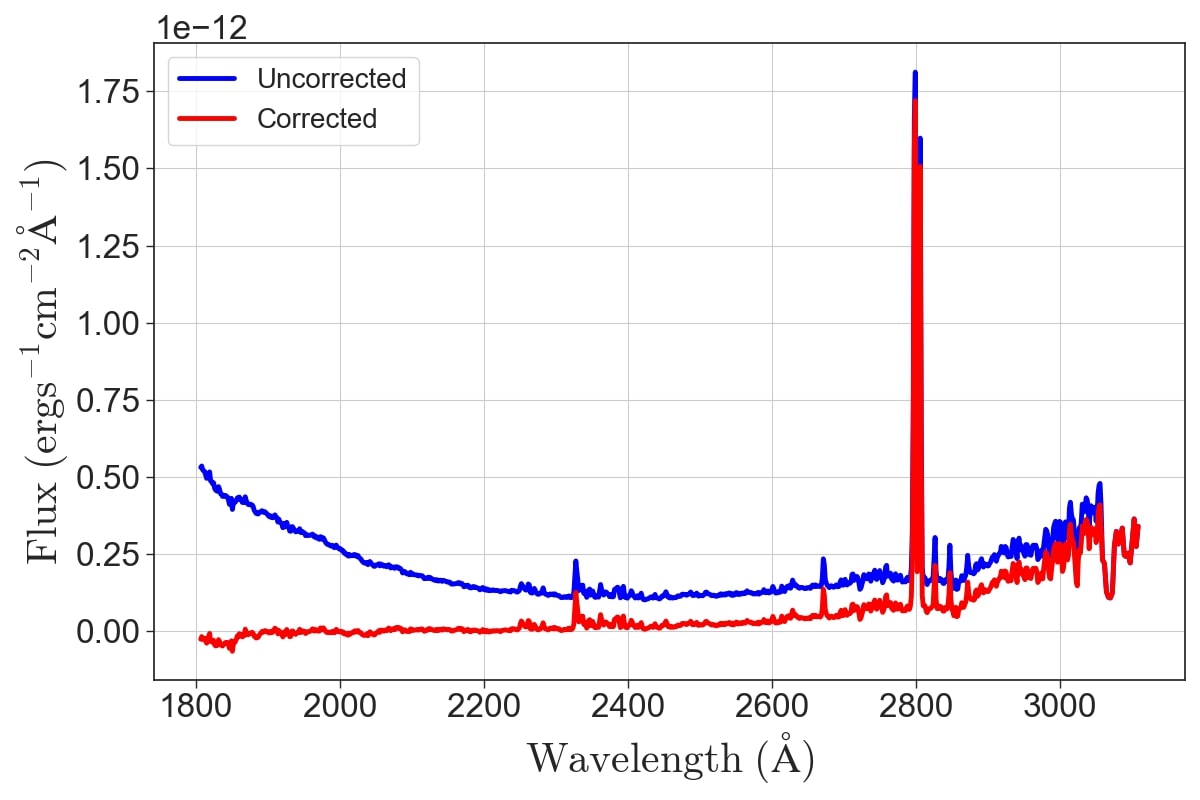}
\caption{The fluxed spectrum of the star HD102212 in the UV region without any scattered light correction (blue) and with scattered light correction (red). It is seen that the spectrum is a little over corrected in the region around 1800\AA}
\label{fig:cor_uncor}
\end{figure}

The scattered light ($S(\lambda)$) is approximated by the formula \citep{2022stis.rept....5W}:
\begin{equation}\label{eqn:scatter_light}
    S(\lambda)=K_0\times(1+0.00104\times(\lambda-2000)) \, ,
\end{equation} where $K_0$ is the scattered light count rate at 2000\AA\ and $\lambda$ is the wavelength. Targets with T$_{eff}<$5000K, $K_0$ is given by the median counts rate around 2000\AA\ (median counts rate for 1950\AA$<\lambda<$2050\AA). Two stars in our list, HD~124547 and HD~200905, are spectroscopic binary stars with T$_{eff}<$5000K. For these two stars, $K_0$ calculated using the average counts rate around 2000\AA\ resulted in over correction of the spectra. After visually inspecting the spectrum for these two stars, the $K_0$ values were modified by hand to mitigate the problem of over correction. Targets with T$_{eff}>$5000K and for which V magnitudes ($m_v$) are available, $K_0$ is given by--
\begin{equation}\label{eqn:k0v}
    K_0 = 426\times10^{-0.4m_v} \, .
\end{equation} But, for some of the targets (with T$_{eff}>$5000K) $m_v$ is not available. For such targets, $K_0$ is given by--
\begin{equation}\label{eqn:k0c}
    K_0 = 1.78\times10^{-7}\times C \, ,
\end{equation} where C is the integrated count rate between 2000\AA\ and 10000\AA. $S(\lambda)$ was then subtracted from overall count at each $\lambda$. Fig.~\ref{fig:cor_uncor} shows an example of scattered light correction applied to the spectrum of HD102212.\\

After applying the above mentioned formula of $S(\lambda)$ for all the 514 stars, 96 stars (T$_{eff}>$5000K) were over corrected and 8 stars (T$_{eff}>$5000K) were under corrected as judged by inspection of the spectra. For these cases, the coefficient values ($426$ in Eqn.~\ref{eqn:k0v} and $1.78\times10^{-7}$ in Eqn.~\ref{eqn:k0c}) was iteratively modified to calculate $K_0$ until the discrepant star fell among its peers in the UV. The updated $K_0$ values were then used to calculate $S(\lambda)$ for those 104 targets.

\subsection{Slit Off-center Correction}\label{subsec:slit_off}

The NGSL targets were observed using the $0\farcs2$ slit. If the target is not placed at the center of the slit, light at the edges of the point spread function (PSF) gets attenuated by the slit edges. Because the STIS instrument is off-axis, the PSF is asymmetric, and the attenuation is wavelength-dependent. To correct for the attenuation effect, we use the attenuation factor ($D_\lambda$) which is given by \citep{2022stis.rept....5W}:
\begin{equation}\label{eqn:slit_off}
    D_\lambda=a+bq+cq^2+dq^3+eq^4+fq^5+gq^6 \, ,
\end{equation} 

where $q = \sqrt{\lambda/4500}$. The coefficients for the above formula at different slit off-center values are given in Table 3 of \cite{2022stis.rept....5W}. The slit off-center value for each of the 514 NGSL spectra was calculated during the defringing process as outlined in $\S$\ref{subsec:defringe}. It is obvious that the slit off-center values for our 514 targets were not matching the exact values given in Table 3 of \cite{2022stis.rept....5W}. The $D_\lambda$ curve (as a function of $\lambda$) for each of our targets was calculated as linearly interpolated curve between two nearest $D_\lambda$ curves (for which coefficients are available from \cite{2022stis.rept....5W}). Once the $D_\lambda$ curve was calculated for each target, the flux of that target was divided by $D_\lambda$ at each $\lambda$ value.

\subsection{Dust}

We compiled interstellar dust extinction data for our 513-star library sample.  \cite{2012AA...538A.143K} gives non-negative A$_V$ values for around 341 stars. A$_V$ for 44 stars are calculated by us \citep[following][]{2018A&A...615A.115K} by matching an observed spectrum with a synthetic spectrum and then fitting a 1-variable extinction law from \cite{1999PASP..111...63F}. The rest of the A$_V$ values are taken from GALExtin website version 1.2 \citep{2021MNRAS.508.1788A} using a three dimensional Galactic extinction model by \cite{2003AA...409..205D}. These A$_V$ values are used to find the E(B-V) values using the following equation: 

\begin{equation}\label{eqn:extinction}
    E(B-V) = \frac{A_V}{3.1}
\end{equation}

The extinction law of \cite{1999PASP..111...63F} was used to correct the spectra to dust-free versions. Possible self-reddening for mass-losing stars was not considered. The $E(B-V)$ values were also used to deredden the observed colors that we use for the analysis below.

\section{Presentation of the Library} \label{sec:data}

\subsection{Archived Spectra}

All 513 spectra have been made available at \url{http://astro.wsu.edu/hststarlib/} and at the Mikulski Archive for Space Telescopes (MAST) as a High Level Science Product via \dataset[10.17909/kmdt-pw63]{\doi{10.17909/kmdt-pw63}} in separate FITS \citep{1981A&AS...44..363W} files. Each FITS file contains 5 extensions, briefly described in Table \ref{tab1:extensions}. 

\begin{table}
\centering
\begin{tabular}{ | P{10em} | P{15em}|} 
  \hline
  \textbf{Extension}& \textbf{Description}\\ 
  \hline
  Primary & Contains no data. The header contains information about basic stellar parameters ([Fe/H], $\mathrm{log} \, g$, etc.) and averaged pointing information. Exposure-level pointing is available from the original MAST archive files. \\ 
  \hline
  Flux Table & Binary table extension with columns for wavelength (in \AA), uncorrected flux, scattered light corrected flux, scattered light \& slit off-center corrected flux, and scattered light, slit off-center \& dust corrected flux (fluxes are in erg/s/cm$^2$/\AA). Flux errors are also included as separate columns. \\ 
  \hline
  Count Rate Table & Binary table extension with columns for wavelength (in \AA), uncorrected count rate, scattered light corrected count rate, scattered light \& slit off-center corrected count rate, and scattered light, slit off-center \& dust corrected count rate. Uncertainties are also included as separate columns. \\ 
  \hline
  Flux Table (Log Scale) & This binary table extension contains the same information as the Flux Table but the wavelengths are spaced on log scale with $\mathrm{log} \, \Delta\lambda$ = 0.0002 \\
  \hline
  Count Rate Table (Log Scale) & This binary table extension contains the same information as the Count Rate Table but the wavelengths are spaced on log scale with $\mathrm{log} \, \Delta\lambda$ = 0.0002 \\
  \hline
\end{tabular}
\caption{Brief description of the FITS file structure.}
\label{tab1:extensions}
\end{table}

Table \ref{tab:meta} summarizes a mixture of astrophysical and reduction-specific metadata for each stellar target. 

\addtolength{\tabcolsep}{-2pt}  
\begin{deluxetable*}{llrrrrrrrrrrrrrrrr}
\tabletypesize{\scriptsize}
\tablecaption{Stellar Metadata}

\tablehead{\colhead{Simbad} & \colhead{Header} & \colhead{$T_{eff}$} & \colhead{log $g$} & \colhead{[Fe/H]} & \colhead{B} & \colhead{V} & \colhead{$\pi$} & \colhead{$(M_V)_0$} & \colhead{dSlit} & \colhead{$v_r$} & \colhead{$K_0$} & \colhead{$A_V$} & \colhead{src} & \colhead{$\rm{Mg2800}$} & \colhead{$\rm{H\alpha}$} & \colhead{$\rm{H\beta}$} & \colhead{Note} \\ 
\colhead{Name} & \colhead{Name} & \colhead{(K)} & \colhead{(dex)} & \colhead{(dex)} & \colhead{(mag)} & \colhead{mag} & \colhead{(mas)} & \colhead{(mag)} & \colhead{(pixel)} & \colhead{(km s$^{-1}$)} & \colhead{(ADU)} & \colhead{(mag)} & \colhead{} & \colhead{(mag)} & \colhead{(mag)} & \colhead{(mag)} & \colhead{} } 

\startdata
HD 60319  &  HD060319  &  5907  &  4.03  &  -0.82  &  9.46  &  \nodata  &  10.99  &  \nodata  &  -0.20  &  -34.1  &  0.2  &  0.08  &  1  & 1.27 & 0.07 & 0.08 & \nodata \\
G 202-65  &  G202-65  &  6656  &  4.25  &  -1.37  &  \nodata  &  \nodata  &  3.88  &  \nodata  &  1.00  &  -245.6  &  0.0  &  0.00  &  1 & 0.61 & 0.12 & 0.15 & \nodata   \\
HD 185351  &  HD185351  &  4921  &  2.95  &  0.01  &  6.11  &  5.17  &  24.22  &  2.00  &  0.80  &  -6.6  &  5.2  &  0.09  &  1 & 0.45 & 0.05 & 0.04 & \nodata   \\
HD 72184  &  HD072184  &  4643  &  2.84  &  0.23  &  7.01  &  \nodata  &  14.55  &  \nodata  &  -0.10  &  16.5  &  2.4  &  0.11  &  1 & 0.15 & -0.01 & 0.03 & \nodata   \\
HD 126614  &  HD126614  &  5453  &  3.87  &  0.53  &  9.66  &  8.79  &  13.65  &  4.41  &  -0.20  &  -32.9  &  0.2  &  0.05  &  1 & 0.67 & 0.03 & 0.08 & \nodata   \\
\enddata

\tablecomments{In this table, B and V are as observed (not dereddened), but $(M_V)_0$ is dereddened. The "src" column is for $V$-band extinction $A_V$: 1 -- \citet{2012AA...538A.143K}; 2 -- Our derivation based on comparison with synthetic templates; or 3 -- \citet{2003AA...409..205D}. \textbf{The "Note" column refers to objects noted in $\S$\ref{subsec:note_obj}: 1 -- noisy; 2 -- possible extraction error; 3 -- chemically peculiar; 4 -- binary that does or may suffer from compositeness; 5 -- photometric variable. } This is a portion of the table, presented to show format and content. The entirety is available online.}
\label{tab:meta}
\end{deluxetable*}
\addtolength{\tabcolsep}{2pt}  

\subsection{Notable objects}\label{subsec:note_obj}

\begin{itemize}
    \item Targeted object Gleise 15B, a late M dwarf in a visual binary system, was not observed. Due to the count rate and spectral shape, it is near certain that its primary (Gleise 15A, GJ 15A, HD 1326, GX And) was observed instead. Our metadata has been updated to reflect this change.
    \item Quite a few chemically peculiar stars were included in the library that practitioners wishing to fit only ``normal'' stars should exclude. HD~319, HD~141851, HD~204041, HD~210111, and HD~210418 are $\lambda$ Bootis stars. HD~18769, HD~41357, HD~41770, HD~67230, HD~78209, HD~95418, HD~109510, HD~111786, HD~140232, HD~141795, and HD~172230 are Am stars. HD~176232 and HD~175640 are Ap/Bp stars. HD~79158, HD~163641, HD~196426, and HD~220575 are Hg-Mn stars. HD~103036 has anomalously-low Mn. CD$-$62~1346 is a carbon-enhanced metal-poor star. HD~183915 and HD~101013 are Ba stars and spectroscopic binaries. HD~30834 and HD~104340 are Ba stars.
    \item HD~54361 is a carbon star and it has very little Mg2800 emission. This might indicate that C-stars have abnormal chromospheres. HD~158377 is also a carbon star and BD+36~3168 is a J-type carbon star.
    \item HD~37202, HD~58343, HD~109387, HD~138749, and HD~142926 are Be stars with strong Balmer emission lines, presumably from a disk. HD~190073 is a Herbig Ae star with similar strong emission. HD~30614 is a blue supergiant star with strong emission for H$\alpha$.
    \item HD~358, HD~15089, HD~18078, HD~34797, HD~72968, HD~78316, HD~108945, HD~112413, HD~137909, HD~176232, HD~201601, and HD~224801 are $\alpha^2$ CVn variable stars, also, broadly, Ap/Bp stars or HgMn stars.
    \item HD~232078 is a metal-poor long-period variable star for which we observe little Mg2800 flux. This star appears in most of the large stellar libraries. It is a probable Mg2800 variable star, since \cite{2007AJ....134.1348D} give a surface flux of $\mathrm{log} \, F=$ 5.17 erg s$^{-1}$ cm$^{-2}$. It has also been observed to have H$\alpha$ emission in the wings of the line \citep{1976ApJ...203L.127C}. We hypothesize that at some phase range of the variability cycle, perhaps during heavy mass loss, the normal chromosphere structure is disrupted.
    \item{Variable stars: HD~173819 is a classical Cepheid variable star. HD~67523 and HD~183324 are $\delta$ Scuti (dwarf Cepheid) variable stars. V* BN VUL, BD~+06~4990, and CD$-$25~9286 are RR Lyrae variable stars. HD~96446 pulsates and is a Bp star. HD~170756 is an RV Tauri variable star.}
    \item{Stars with some degree of binary compositeness include HD~41357, HD~69083, HD~78362, HD~79469, HD~106516, HD~164402, HD~166208, HD~187879, HD~193495, HD~210111. In our spectra, extra UV light from a companion can be seen in HD~26630, HD~124547, and HD~200905.}
    \item{HD~149382 is a hot subdwarf (sdB) star. The origin of these stars is not perfectly clear, but they are highly evolved.}
    \item{HD~1638 and LHS~10 have noisy spectra. For purposes of repeatability, we did not pursue alternative spectral extraction methods, but we note that \textit{stistools.x1d}'s extractions for at least G~63-26, G~115-58, G~169-28, G~192-43, G~196-48, and BD~+66~268 are probably incorrect.}
    
\end{itemize}

The list above points out a problem for this library among AB main sequence stars: chemically-peculiar stars comfortably outnumber chemically-normal stars. 

\section{The Mg II 2800 Feature and Chromospheric Activity} \label{sec:mg}

In this section, we explore the chromospheric activity of the 513 NGSL stars after full reduction, including extinction corrections. 
\cite{1957ApJ...125..661W} showed that the absolute visual magnitudes (M$_V$) of late-type stars correlate linearly with logarithm of H \& K emission line width of CaII (the Wilson-Bappu effect) and Mg2800 h \& k share this behavior \citep{1999A&A...343..222E, 2001A&A...374.1085C}. However, because our spectra are low resolution we could not reliably compute an analogous width for the twin MgII 2800 emission lines. We therefore measure overall strength only. 

\begin{figure}[!ht]
\centering
\includegraphics[width=8.6cm]{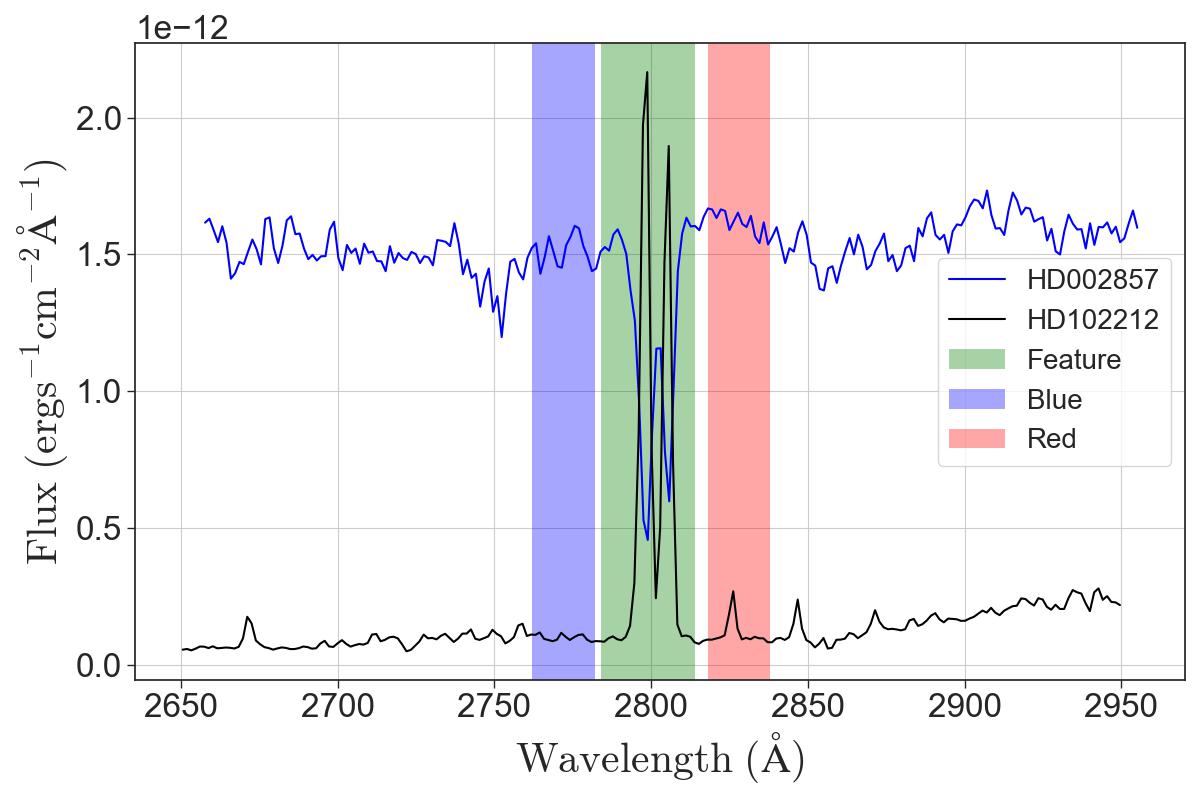}
\caption{Spectra of two stars, HD002857 (T$_{eff}=$7607K, as blue line) and HD102212 (T$_{eff}=$3738K, as black line), showing feature band (green), blue side band (blue), and red side band (red) for our calculation of Mg2800 feature strength. The flux for HD002857 is scaled up by a factor of 10 for visual prominence. The hotter star (HD002857) and the cooler star (HD102212) has Mg2800 in absorption and emission respectively.}
\label{fig:mg_feature}
\end{figure}

 To summarize the strength of MgII 2800 emission, we adopt an equivalent width style index ($\rm{Mg2800}$):
 \begin{equation}\label{eqn:mag}
    \rm{Mg2800} = -2.5\times log_{10}\frac{\int F_\lambda^i\ d\lambda}{\int F_\lambda^c\ d\lambda} \, ,
\end{equation} 

where $F_\lambda^i$ is the observed flux within the spectral feature band and $F_\lambda^c$ is the expected flux without the spectral feature within the same band. We approximate $F_\lambda^c$ by defining a pseudo-continuum from side bands. A line is drawn between the central wavelengths and average flux values of the two sidebands. The Mg2800 central feature band is defined as wavelengths between [2784\AA, 2814\AA]. The blue side band is [2762\AA, 2782\AA] and the red one is [2818\AA, 2838\AA]. These definitions of feature and side bands are adopted from \cite{1990ApJ...364..272F}. Fig.~\ref{fig:mg_feature} shows the adopted definitions for Mg2800 central feature band and the two side bands (blue and red).

\begin{figure}[!ht]
\centering
\includegraphics[width=8.6cm]{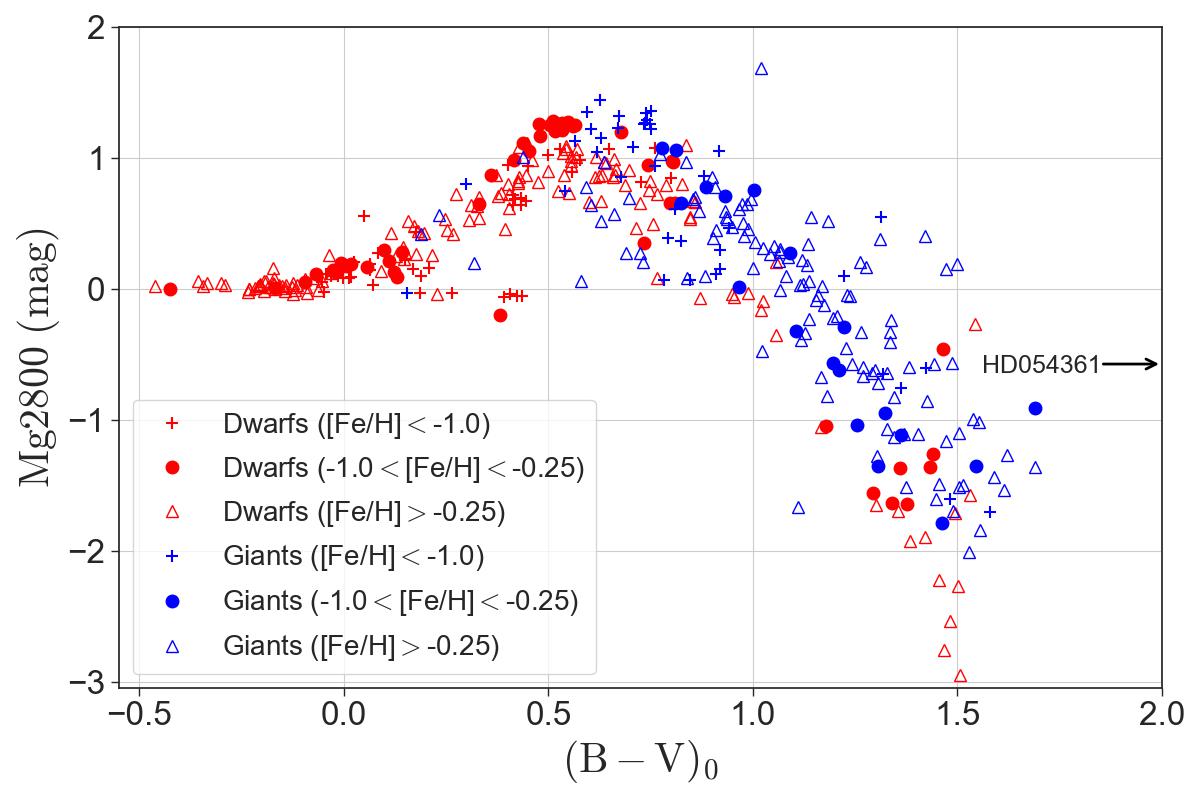}
\caption{Mg2800 versus (B-V)$_0$. Dwarfs (red) and giants (blue) are given different symbol types to denote metallicity groups: metal-poor (crosses), intermediate (filled circles), and metal-rich (non-filled triangles). Carbon star HD~54361 lies outside the plot limit and its position is indicated by a black arrow to the right.}
\label{fig:mag_col_Mg}
\end{figure} 

We keep the units (magnitudes) adopted by \cite{1990ApJ...364..272F}. A negative index value signifies net emission and a positive value signifies absorption. Fig.~\ref{fig:mag_col_Mg} displays Mg2800 as a function of dereddened color for the library stars. Hot stars have negligible Mg2800 absorption. We also note that, although the sample contains some strongly-active Be stars, these stars show no anomalous Mg2800 absorption or emission. Mg2800 absorption increases from A0 stars to sunlike stars [$(B-V)_0=0.65$] and declines thereafter. In cool stars, both giants and dwarfs, chromospheric Mg2800 emission overtakes photospheric absorption at $(B-V)_0 \approx 1$ and dominates for cooler stars. Fig.~\ref{fig:mag_col_Mg} agrees well with Fig.~5c of \cite{1990ApJ...364..272F}. 

\begin{figure}
\centering
\includegraphics[width=8.6cm]{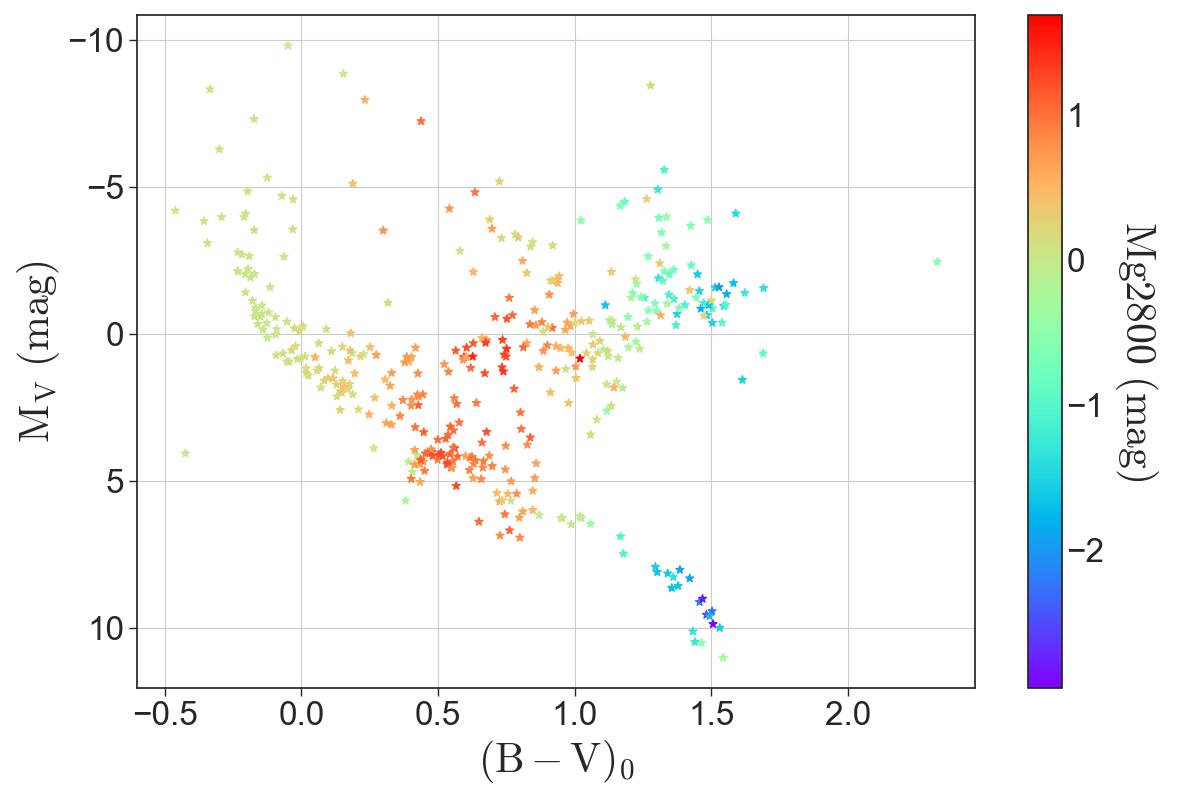}
\caption{CMD for all 513 NGSL stars. The color bar shows Mg2800 strength. Stars warmer than $(B-V)_0 = 0$ or fainter than $M_V = 3.0$ were considered dwarfs. For $(B-V)_0 > 0$, any star with $M_V>6.25\times(B-V)_0-2.5$ is considered a dwarf whereas $M_V<6.25\times(B-V)_0-2.5$  is considered a giant. For dwarfs, Mg II emission fills in the absorption redder than $B-V=0.9$, whereas emission begins to dominate for giants at $B-V=1.2$.}
\label{fig:cmd_Mg}
\end{figure}

For the plots herein, the distinction between giants and dwarfs is approximated via the color-magnitude diagram (CMD) as shown in Fig.~\ref{fig:cmd_Mg}. Stars warmer than $(B-V)_0 = 0$ or fainter than $M_V = 3.0$ were simply considered dwarfs regardless of their spectral type. For $(B-V)_0 > 0$, any star with $M_V>6.25\times(B-V)_0-2.5$ is considered a dwarf whereas $M_V<6.25\times(B-V)_0-2.5$  is considered a giant. 

The Fig.~\ref{fig:cmd_Mg} CMD is color-coded by Mg2800 value. The verticality of the color bands shows again that both cool dwarfs and cool giants have similar Mg2800. Their chromospheres are similar by this measure despite vastly different size scales ($\sim0.1 R_\odot$ versus $\sim100 R_\odot$). The emission gradually changes to absorption for warm stars and declines to near zero for hot stars. Note that some distant stars may have extra Mg2800 absorption due to warm interstellar material along the line of sight. 

Even given the intentional diversity in sample selection, outliers are relatively few. One is G9 giant HD~222093, at $(B-V)_0 \approx 1$ and M$_V \approx 1$ in Fig.~\ref{fig:cmd_Mg}. It has a high value for Mg2800 absorption, signified by the red color in Fig. \ref{fig:cmd_Mg}. The star's spectrum shows emission peaks at the core of a broad absorption feature at 2800\AA, normal for a star whose absorption competes with emission at $(B-V)_0 \approx 1$, but this star's emission is weak. HD~222093 also shows up in Fig.~\ref{fig:mag_col_Mg} as the sole star with the highest Mg2800 absorption at $(B-V)_0 \approx 1$. 

\begin{figure}
\centering
\includegraphics[width=8.6cm]{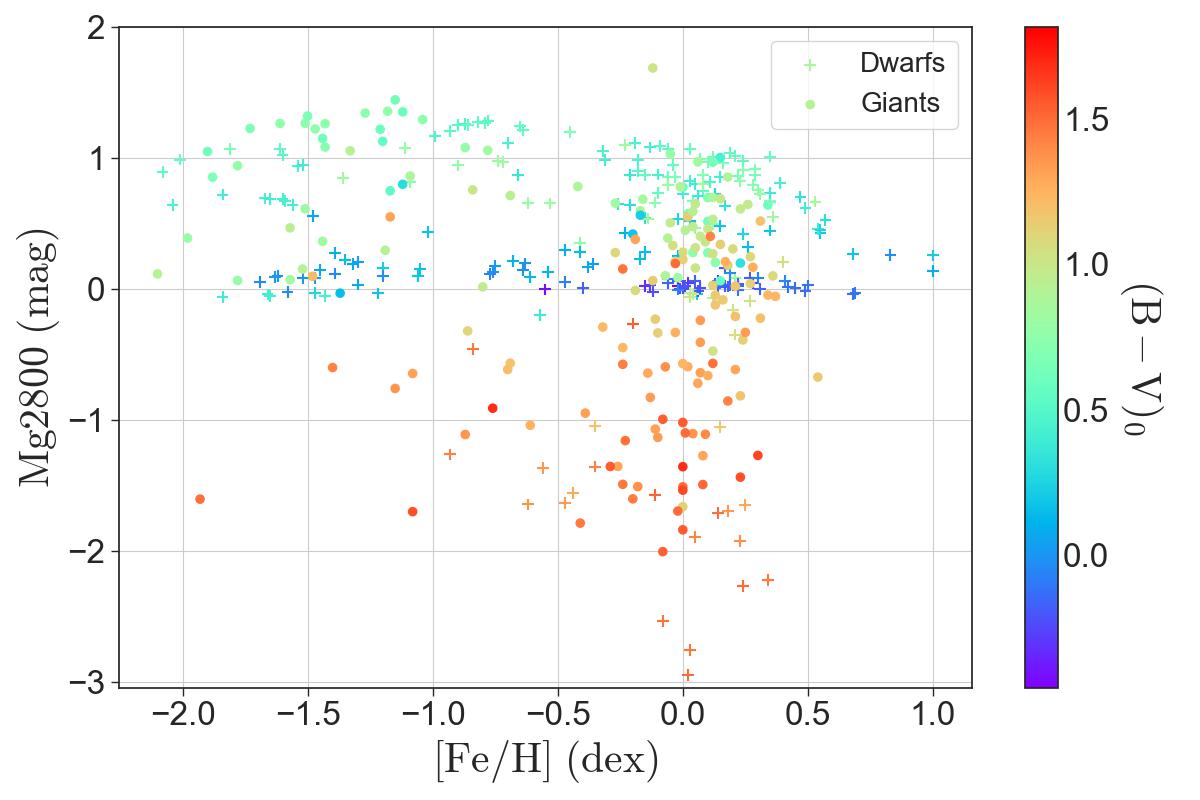}
\caption{Mg2800 as a function of [Fe/H]. The color bar codes (B-V)$_0$ and the symbol type distinguishes dwarfs (crosses) and giants (circles).}
\label{fig:met_logg_mag_Mg}
\end{figure}

Fig.~\ref{fig:met_logg_mag_Mg} plots Mg2800 vs. metallicity, color-coded by $(B-V)_0$. It is clear from this figure that no strong correlation exists between these two quantities in any color regime, particularly for cools. An anticorrelation among cool stars might have been expected from the Ca II H \& K results of \cite{1997A&A...326.1143H} who found that metal poor stars are activity deficient, but we see no such trend. \cite{1997ApJ...480L..47P} reports that chromospheric characteristics do not have any metallicity dependence.

\begin{figure}
\centering
\includegraphics[width=8.6cm]{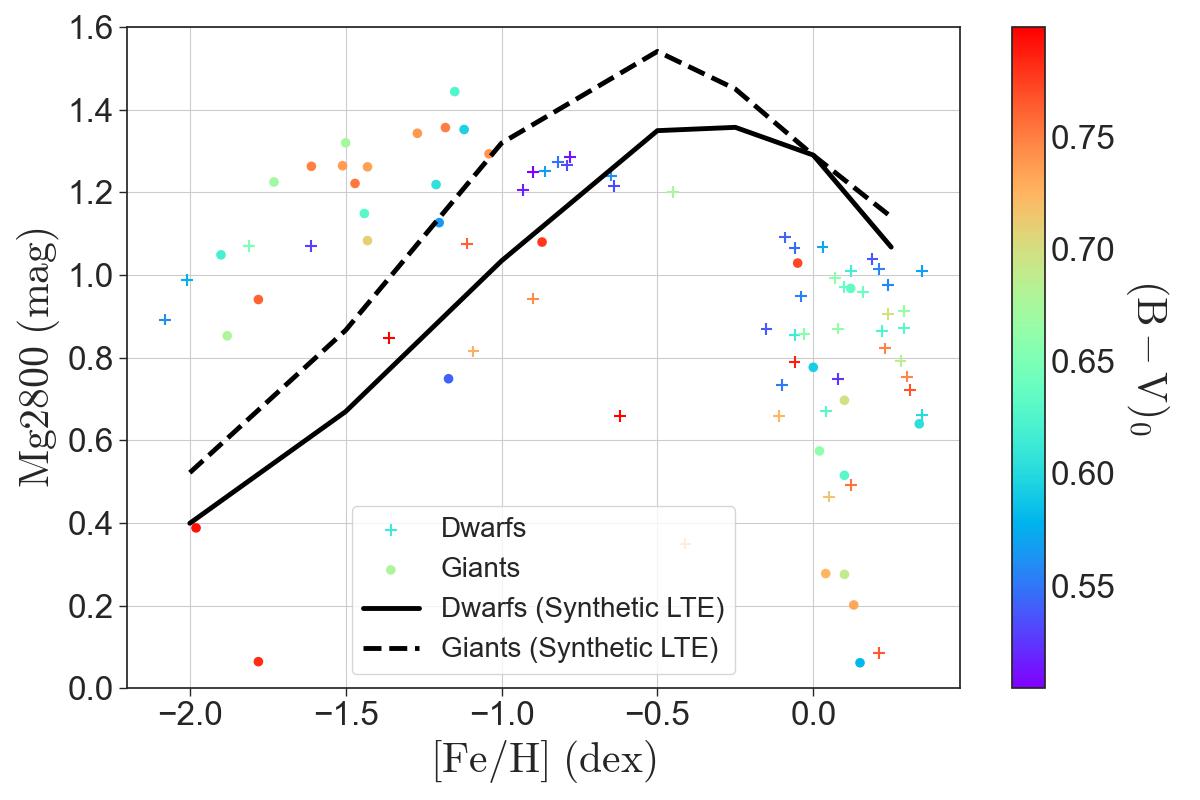}
\caption{Mg2800 as a function of [Fe/H] for a narrowed color range of 0.5$<$(B-V)$_0<$0.8. Dwarfs (crosses) and giants (circles) are color coded by (B-V)$_0$. Black lines indicate Mg2800 from synthetic LTE spectra for dwarfs ($T_{\rm eff}$=5770K, $\mathrm{log} \, g$=4.5, solid) and giants ($T_{\rm eff}$=5770K, $\mathrm{log} \, g$=1.5, dashed). }
\label{fig:fe_h_Mg_dwarfs}
\end{figure}

A subtle declining trend among medium-temperature stars in Fig.~\ref{fig:met_logg_mag_Mg} deserves a note and an additional figure, namely Fig.~\ref{fig:fe_h_Mg_dwarfs}, which restricts the color range to be near solar (0.5$<$(B-V)$_0<$0.8). Because these are positive values of Mg2800, indicating absorption, one might expect a monotonic increase of Mg2800 with [Fe/H]. Mg2800 absorption does increase for metal poor stars ($-2 <$ [Fe/H] $< -1$) but then the index value saturates and falls for metal rich objects. Although \cite{2011arXiv1107.5325L} reports not much of a correlation with Ca II lower envelope and metallicity, but a decreasing trend can be seen in \cite{2022A&A...658A..57M} for [Fe/H]$>$-1. With the help of synthetic spectra, two sequences of which are also plotted in Fig.~\ref{fig:fe_h_Mg_dwarfs}, the reason appears to be a simple curve of growth argument. Mg2800 is a resonance feature that scales approximately as the abundance of the Mg II ion. It reaches full depth at [Fe/H] $\sim -1$, but the flanking (in wavelength) absorption features from a plethora of atomic species are still weak. From [Fe/H] $\sim -1$ and higher, these weak features will grow faster than the central Mg II absorption pair. As the pseudocontinuum drops, the Mg2800 index drops. Parenthetically, the relatively poor agreement of synthetic spectra and observed spectra in Fig.~\ref{fig:fe_h_Mg_dwarfs} should be no surprise. The UV spectrum is crowded, its lines have not received as much attention as optical ones, and for warm and cool stars the wavelength regime is on the blue side of the blackbody curve, exposing defects in the upper layers of the model atmosphere due to the absence of backwarming.

H$\alpha$ emission is a separate indicator of stellar chromospheric activity \citep{1995A&A...294..165M, 2007A&A...469..309C, 2014A&A...566A..66G} and also magnetic flare activity. An index for the H$\alpha$ feature is calculated using the passband definitions of \citet{1998ApJ...496..808C} but here we convert it to magnitude units (Eqn.~\ref{eqn:mag}). The spectral feature band is [6548\AA, 6578\AA] and the blue pseudocontinuum is [6420\AA, 6455\AA] and the red pseudocontinuum is [6600\AA, 6640\AA]. 

\begin{figure}
\centering
\includegraphics[width=8.6cm]{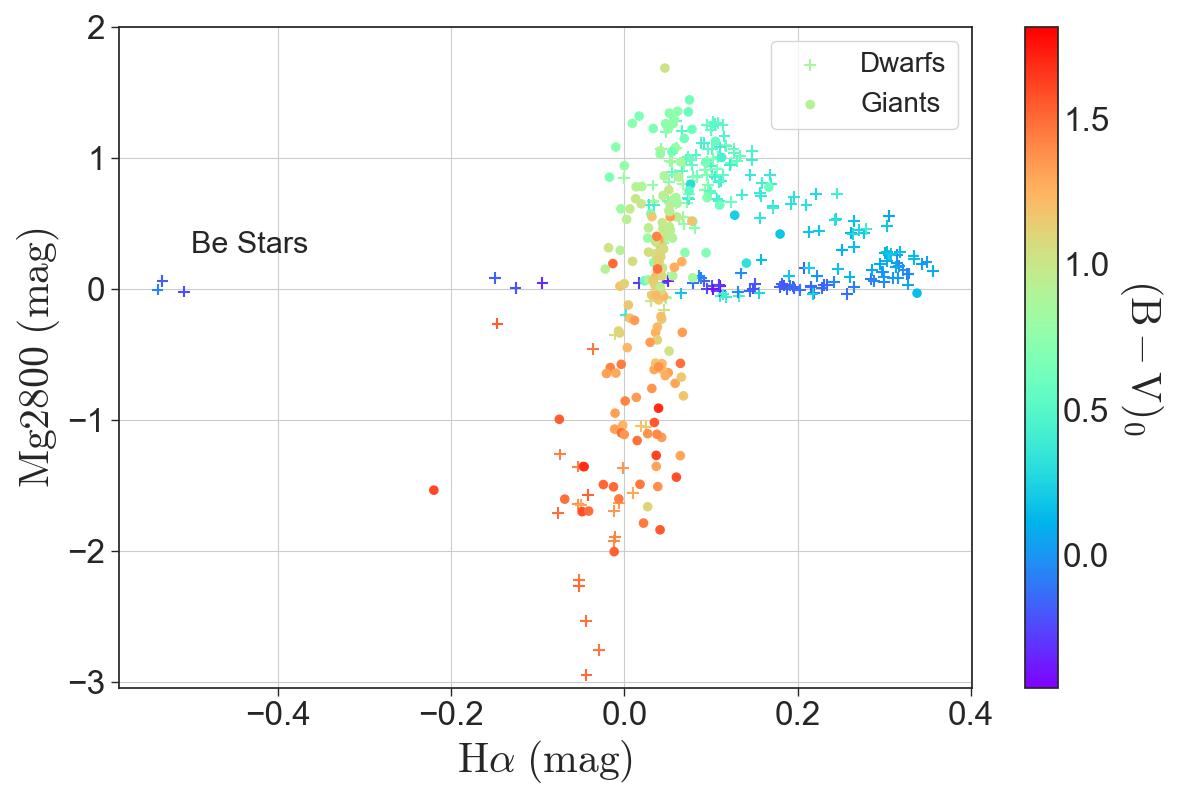}
\caption{Mg2800 is plotted against H$\alpha$ for dwarfs (crosses) and giants (circles). The points are color coded by (B-V)$_0$. Three stars to the extreme left of the figure are all Be stars: HD~37202, HD~109387, and HD~190073.}
\label{fig:H_Mg_mags}
\end{figure}

Mg2800 and H$\alpha$ are plotted against each other in Fig.~\ref{fig:H_Mg_mags}. The strongest H$\alpha$ emitters are Be stars, generally assumed to be young stars with disks \citep{2009ssc..book.....G}. We might also expect to catch some flaring M dwarfs, but apparently none of the M dwarfs were observed during outbursts, as we see no cool dwarfs scattering to negative H$\alpha$ values. The ``triangle'' in the positive-positive quadrant arises because peak H$\alpha$ absorption occurs among hotter stars than peak Mg2800 absorption. Among cool stars with negative Mg2800, the mild correlation is due to expected H$\alpha$ index absorption behavior from species unrelated to H$\alpha$ itself, such as TiO \citep[e.g.][]{2004ApJS..152..251V}. That is, it is a consequence of the strong Mg2800-temperature anticorrelation in cool stars, and does not imply H$\alpha$ emission at all. The chromospheric emission has been studied using Ca II H \& K and H$\alpha$ as activity indicators. Some of these studies find a good correlation between these two indicators \citep[please see the review by][and references there in]{2017ARA&A..55..159L} whereas some studies find strong temporal anti correlation between these two indices in certain percentage of their selected stars \citep{2007A&A...469..309C, 2014A&A...566A..66G}. \cite{2022A&A...658A..57M} suggests that this type of behaviour needs to be explained by considering contributions from plages and filaments.

Two stars lie at anomalously-negative H$\alpha$ values. They are: HD~126327 (giant) and GL~109 (dwarf). Presumably, HST serendipitously observed these objects during flare events.

The correlation between Ca II H \& K core emission strength (a third stellar activity indicator) and H$\alpha$ emission is also well studied. Some authors report a positive correlation between the two \citep{1991A&A...251..199P, 1995A&A...294..165M}, some a lack of correlation, and some a negative correlation \citep{2007A&A...469..309C, 2011A&A...534A..30G}. Our Mg2800 results shed little insight into this uncertain area.

\section{Discussion and Conclusion} \label{sec:conclusion}

This paper presents a new reduction of the Next Generation (HST/STIS low resolution) Spectral Library that includes updated flux calibration work, updated scattered light corrections, and an increase in sample size (345 to 513) due to inclusion of stars from run GO13776. This increases the parameter space coverage in $\mathrm{log} \, g$, $T_{\rm eff}$ and [Fe/H] (Figs.~\ref{fig:NGSL} and \ref{fig:met}).

After correction for interstellar extinction, the spectra were used to explore the chromospheric activity of stars using the Mg II 2800 h + k feature and H$\alpha$ as likely indicators.

Against color, there is a gradual change of sign of Mg2800 from positive to negative (signifying absorption to emission transition) for both dwarfs and giants within 0.5$<$(B-V)$_0<$1.5. From Fig.~\ref{fig:mag_col_Mg} it is evident that the transition happens at (B-V)$_0$=1.0 or spectral class K3 for dwarfs, and (B-V)$_0$=1.12 or spectral class K4-K5 for giants. The color calibration of \cite{2011ApJS..193....1W} indicates that we expect dwarfs to have $B-V$ bluer than giants by about 0.1 mag, so this crossover happens at about the same $T_{\rm eff}$ for both dwarfs and giants. Largely, this result is consistent with results from \cite{1975PASP...87..289G} where it was shown that Mg II 2800 feature starts dominating in emission in K2 and later-type stars. The photospheric absorption gives way to strong chromospheric emission as the temperature drops. Temperature is the emphatic controlling parameter of Mg2800 emission; the cooler the star, the stronger the emission. [Fe/H] and log $g$ have little influence on Mg2800, and we see no evidence of flare behavior.

\begin{figure}
\centering
\includegraphics[width=8.6cm]{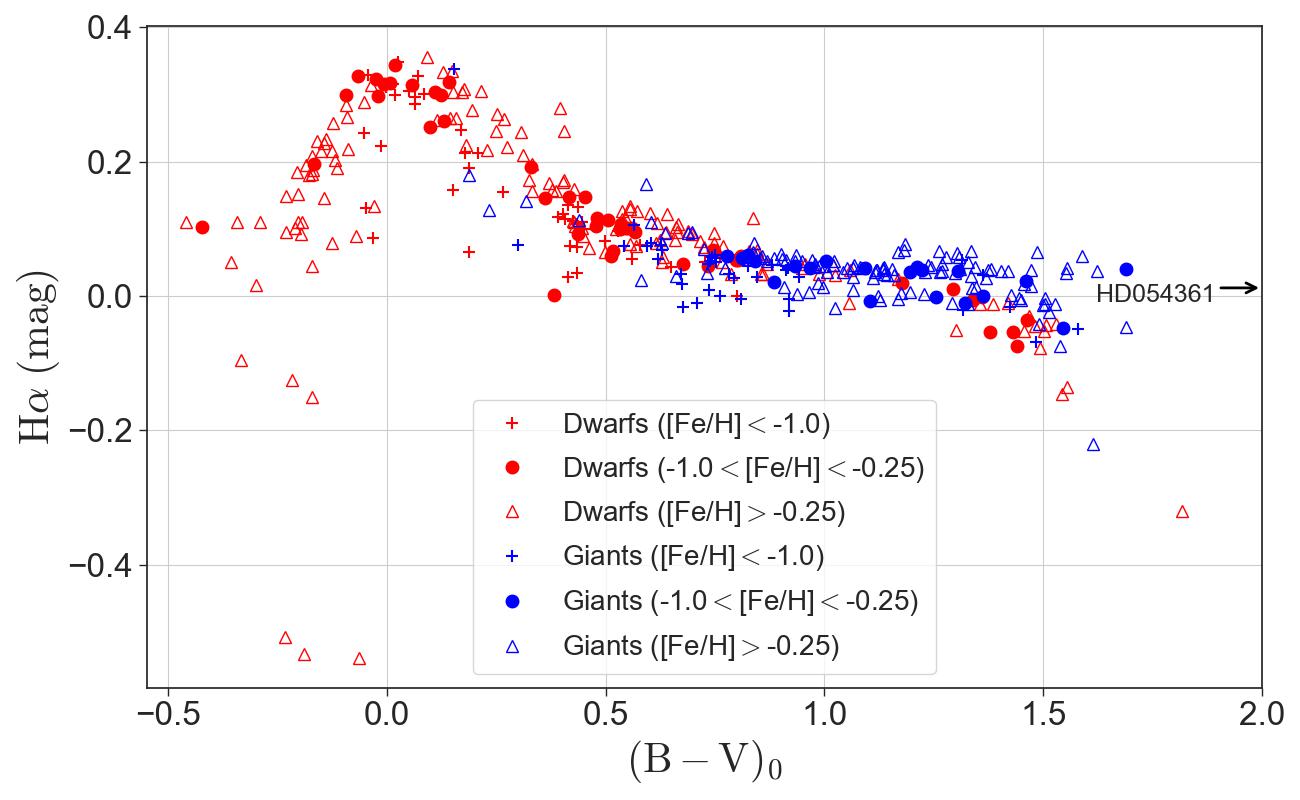}
\caption{H$\alpha$ as a function of $(B-V)_0$ for dwarfs (red) and giants (blue) are shown, segregated by metal-poor (crosses), intermediate (filled circles), and metal-rich (non-filled triangles) status. Be stars scatter to negative values for hot stars with $(B-V)_0 < 0$. Any star caught during a flare event should also scatter toward negative index values. Four stars (3 dwarfs and 1 giant) with H$\alpha<-0.15$ and $(B-V)_0>1.5$ are thought to be flaring: GJ~551, GJ~876, and GL~109 are dwarfs while HD~126327 is a giant. Noise prevents reliable measurement of Mg2800 in GJ~551 (Proxima Centauri) and GJ~876. Therefore, these stars do not appear in figures that illustrate Mg2800. Carbon star HD~54361 lies outside the plot limit and its position is indicated by a black arrow to the right.}
\label{fig:h_alp}
\end{figure}

\begin{figure}
\centering
\includegraphics[width=8.6cm]{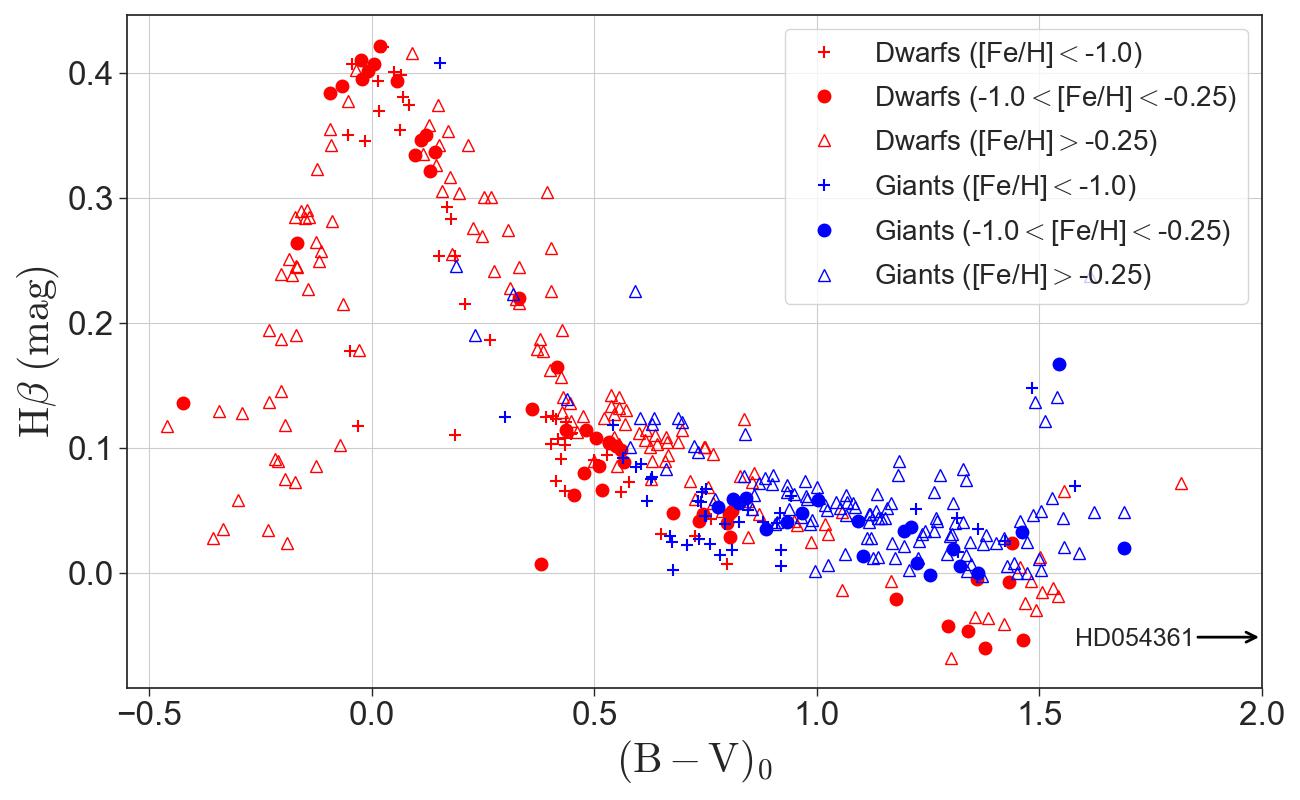}
\caption{H$\beta$ as a function of $(B-V)_0$ for dwarfs (red) and giants (blue), segregated by metal-poor (crosses), intermediate (filled circles), and metal-rich (non-filled triangles) status. H$\beta$ is less sensitive to emission than H$\alpha$. Carbon star HD~54361 lies outside the plot limit and its position is indicated by a black arrow to the right.}
\label{fig:h_beta}
\end{figure}

\begin{figure}
\centering
\includegraphics[width=8.6cm]{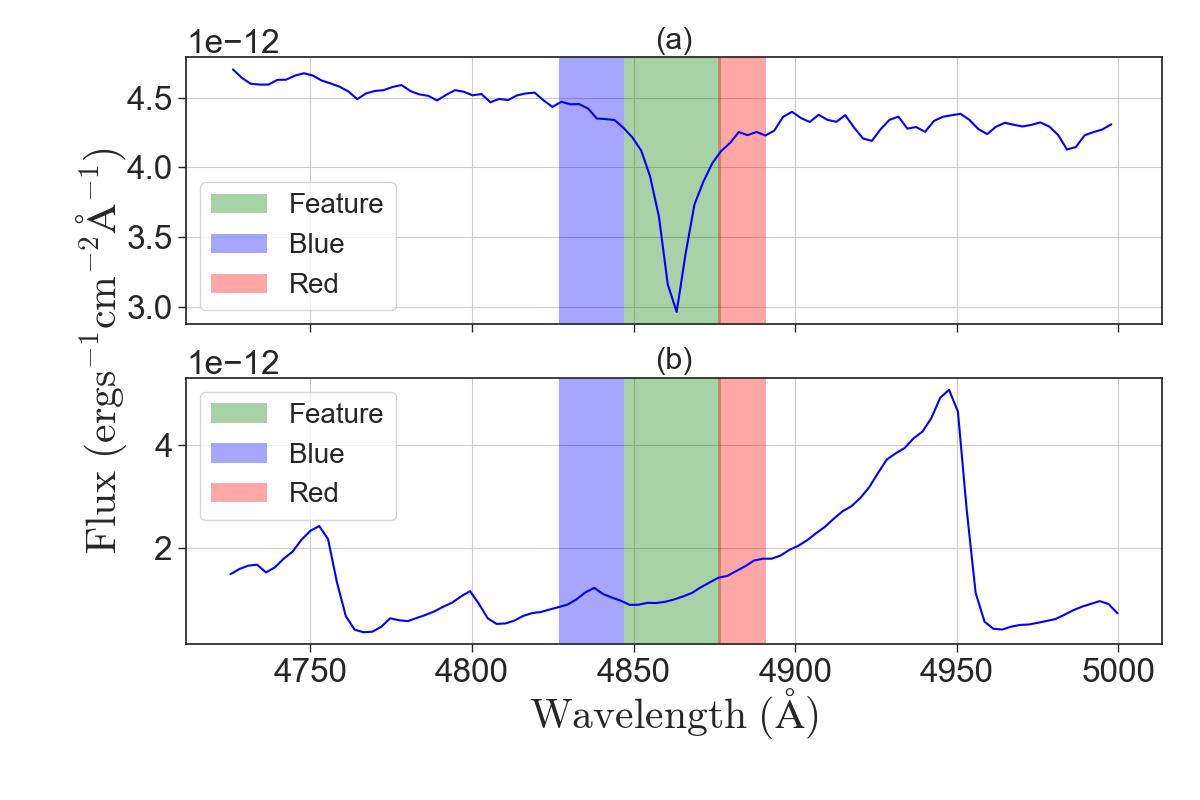}
\caption{Spectra of two stars, HD221377 (T$_{eff}=$6399K, in upper panel (a)) and HD126327 (T$_{eff}=$2874K, in lower panel (b)), showing feature band (green), blue side band (blue), and red side band (red) for our calculation of H$\beta$ feature strength. For the hotter star (HD221377, upper panel (a)), we see clear absorption for H$\beta$ whereas the lower panel (b) shows no apparent activity for H$\beta$ for the cooler star (HD126327). }
\label{fig:hb_feature}
\end{figure}

We chart basic H$\alpha$ and H$\beta$ behavior in Figs.~\ref{fig:h_alp} and \ref{fig:h_beta} respectively. The peaks are the deep absorptions in A stars, and strongly negative values indicate that emission has overshadowed absorption. Fig.~\ref{fig:h_alp} shows four stars with mild flares in progress: GJ~551, GJ~876, and GL~109 are dwarfs while HD~126327 is a giant. GJ~551 is Proxima Centauri and it shows up as a flaring dwarf in a 20 seconds cadence Transiting Exoplanet Survey Satellite (TESS) monitoring campaign \citep{2022ApJ...926..204H}. Evidence for flares in GJ~876 is reported in \cite{2019ApJ...871L..26F}. GL~109 is listed as an eruptive variable in SIMBAD and categorized as UC Cet-type flare star \citep{1999A&AS..139..555G}. HD~126327 is the only cool giant that seems to be flaring. Prominent TiO band absorption affects the coolest stars. Cool giants saturate at $B-V\approx 1.65$ \citep{2011ApJS..193....1W} but especially H$\beta$ continues to increase, not because of actual H$\beta$ absorption, but because of the increasing influence of TiO features. Fig.~\ref{fig:hb_feature} shows the feature and side bands used in calculation of H$\beta$ index strength. The hotter star (upper panel (a)) shows clear H$\beta$ absorption. But for the cooler star (lower panel (b)), the values obtained for the H$\beta$ strengths are mostly from TiO features. The hot dwarfs with H$\alpha$ magnitudes less than -0.1 are Be stars.

The Mg II 2800 line emission in UV is a major probe for chromospheric radiative loss\citep{1978ApJ...220..619L}. From Fig.~\ref{fig:mag_col_Mg} it is evident that there is scatter in Mg II 2800 line strength for a given temperature, but the character of that scatter might be astrophysical. Various studies have suggested the existence of a `basal' flux level for Mg II 2800 that might indicate the level of an ongoing, persistent mechanism (acoustic waves are often cited) that can be supplemented by a more variable heating mechanism (such as magnetohydrodynamic shocks) that adds Mg emission to some stars but not others \citep{1987A&A...172..111S, 1994A&A...281..855S, 10.1111/j.1365-2966.2011.18421.x}.

Recast in terms of the Mg II $\lambda$2800 flux emerging from the star's surface ($F_\lambda$), the above authors find a `basal level' that increases with temperature. In order to confirm this, we select NGSL stars with $T_{\rm eff} <$ 5000K and recast their emission line strengths as emergent fluxes as in \cite{10.1111/j.1365-2966.2011.18421.x}. The scheme follows  \cite{1982A&A...110...30O}, but extended to account for interstellar extinction. Oranje et al. noted that
\begin{eqnarray}
\frac{F_\lambda}{f_\lambda} = \frac{F_{bol}}{f_{bol}} \, ,
\end{eqnarray}

where $F_\lambda$ is the star's outbound surface flux (erg cm$^{-2}$ s$^{-1}$) at some wavelength. For us, this wavelength is 2800\AA , and it is chromospheric in origin. The lower case $f_\lambda$ is then the flux received at earth. The right hand side are the bolometric versions. This equation is only good in the limit of zero extinction. Extinction at wavelength $\lambda$ (A$_\lambda$) is defined by:
\begin{eqnarray}
A_\lambda = -2.5 {\rm log} \ \frac{f_\lambda}{f_{0,\lambda}} \, ,
\end{eqnarray}
where $f_{0,\lambda}$ is the extinction corrected version of $f_\lambda$. This equation can be inverted to
\begin{eqnarray}
f_\lambda = f_{0,\lambda} 10^{-0.4A_\lambda} = f_{0,\lambda}\  g(A_\lambda) \ ,
\end{eqnarray}
where the function $g(A_\lambda)$ is shorthand we introduce. By convention, $A_\lambda$ is positive and thus, $f_\lambda$ is always less than $f_{0,\lambda}$ and $0 < g(A_\lambda) \leq 1$. Besides $g(A_\lambda)$, we also invent $h(A_{bol})$ to represent the extinction in bolometric quantities, which is more complicated to produce (it requires the integration of the dust-attenuated flux over all wavelengths and thus depends on the spectral type of the target star). Putting everything together:
\begin{eqnarray}
F_\lambda = f_{0,\lambda} \frac{F_{bol}}{f_{0,bol}} \frac{g(A_\lambda)}{h(A_{bol})}
\end{eqnarray}
\noindent This can be rephrased in terms of $T_{\rm eff}$ by noting that:
\begin{eqnarray}
F_{bol} = \sigma T_{\rm eff}^4 \, ,
\end{eqnarray}

where $\sigma$ is the Stephan-Boltzmann constant and
\begin{eqnarray}
f_{0,bol} = B 10^{ -0.4(V + BC_V) } \, ,
\end{eqnarray}

where $B$ is a zeropoint adjustment between physical units and the astronomical magnitude scale, $V$ is the apparent magnitude in V-band, and $BC_V$ is the bolometric correction for V-band. The $B$ value is obtained by noting that, $f_{bol,\odot}$=1361 Wm$^{-2}$, V$_\odot$=-26.76 \citep{2018ApJS..236...47W}, and BC$_{V,\odot}$=0.09 \citep{2003AJ....126..778V}. 

\begin{figure}
\centering
\includegraphics[width=8.6cm]{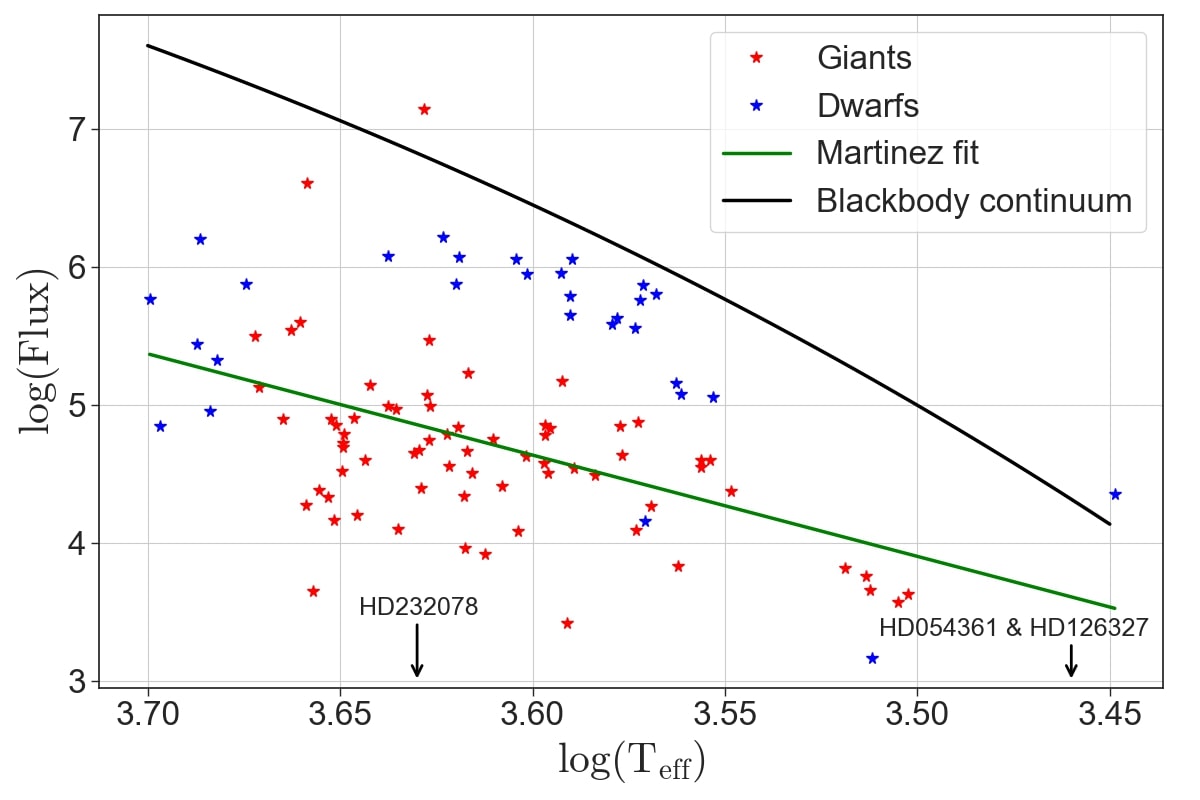}
\caption{Inferred surface flux from Mg II 2800 (log$_{10}$, cgs units) as a function of $T_{\rm eff}$ for both giants (red) and dwarfs (blue) with $T_{\rm eff}<$ 5000 K. The green line is the ``basal flux'' from \cite{10.1111/j.1365-2966.2011.18421.x}. Three stars with $\mathrm{log} \, (Flux)<$3.0 (HD~54361, HD~126327, and HD~232078) are below the plot limits. The downward black arrows show $\mathrm{log10} \, (T_{\rm eff})$ for them. For comparison, the blackbody emergent flux integrated over the Mg2800 central passband (black line) is shown.}
\label{fig:basal_Mg}
\end{figure}

\begin{figure}
\centering
\includegraphics[width=8.6cm]{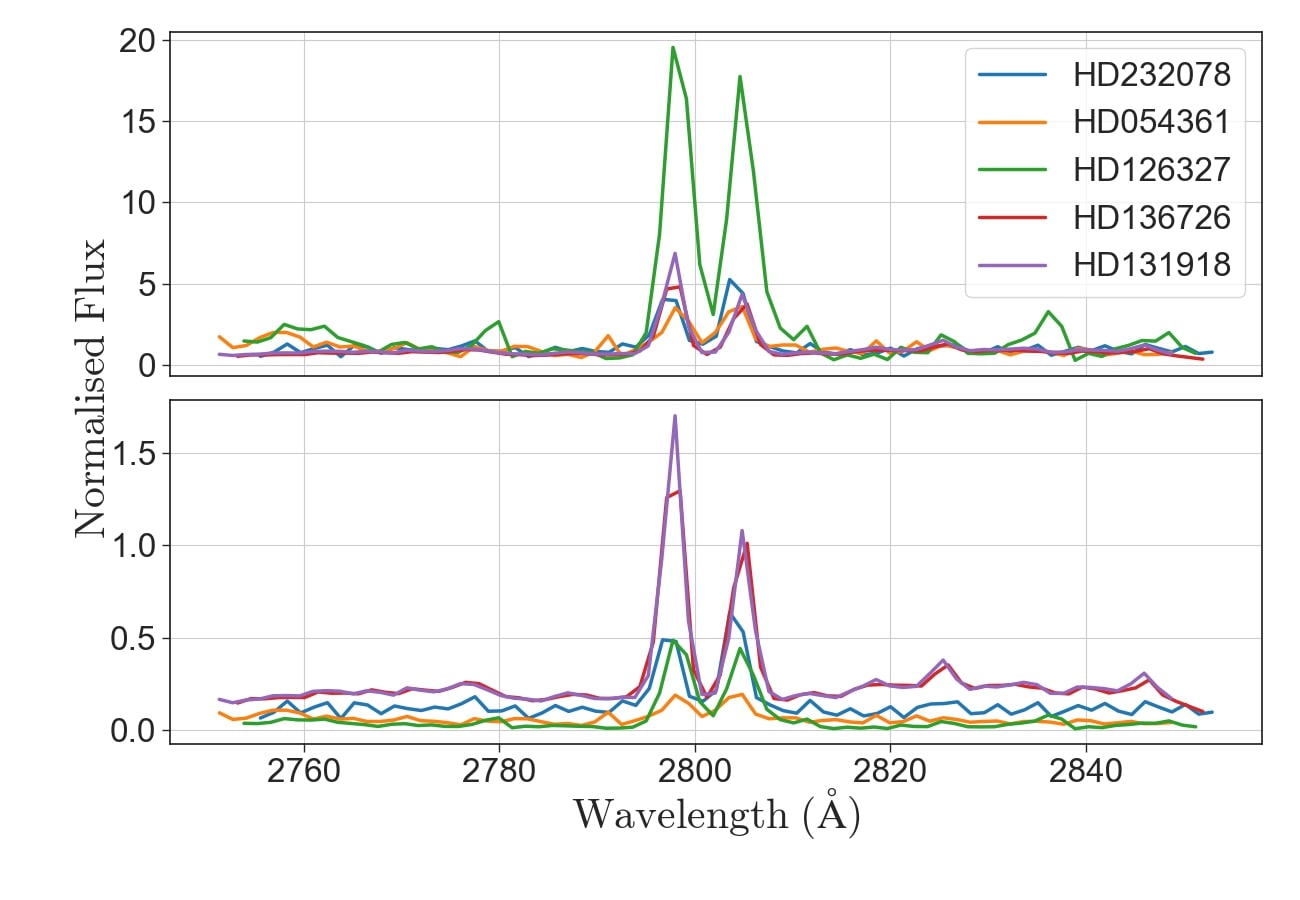}
\caption{Spectra of 5 stars are shown in the $\lambda$2800 region. TOP: Fluxed spectra are normalized at 2820\AA. BOTTOM: Fluxed spectra are normalized such that the continuum-subtracted emission scales as the surface-emergent emission $F_\lambda$ derived for Fig. \ref{fig:basal_Mg}. ``Normal'' HD~136726 and HD~131918 lie near the green line in Fig.~\ref{fig:basal_Mg} and the remaining three stars are low outliers. HD~232078 and carbon star HD~54361 lie outside the plot limits in Fig.~\ref{fig:basal_Mg} and HD~126327 was caught during a flare event (Fig. \ref{fig:H_Mg_mags}).}
\label{fig:low_high_basal}
\end{figure}

Known $T_{\rm eff}$, [Fe/H], and $\mathrm{log} \, g$ values for each star were used to interpolate a low resolution synthetic flux from \cite{1994ApJS...95..107W}. We applied a \cite{1999PASP..111...63F} cubic spline extinction curve to this synthetic flux, then integrated (with and without extinction) to find $h(A_{bol})$. For the bolometric correction, we used the \cite{2011ApJS..193....1W} calibration, which also requires $T$, $\mathrm{log} \, g$, and [Fe/H]. We used these values and our Eqn.~\ref{eqn:extinction} to get A$_{2800}$. The quantity $f_{0,bol}$ was calculated by integrating the flux over index band for Mg II 2800. A linear pseudocontinuum calculated from the Mg II 2800 passbands was subtracted before the integration. 

Fig.~\ref{fig:basal_Mg} shows the dependence of $F_\lambda$ as a function of $T_{\rm eff}$ in a log-log scale. Thus transformed to surface-emergent flux, cool dwarfs are seen to emit an order of magnitude more Mg2800 flux per unit surface area, with two notable low-lying objects. As for giants, a number of cool giants have lower flux values than the basal line given by \cite{10.1111/j.1365-2966.2011.18421.x} (solid green line in Fig.~\ref{fig:basal_Mg}). One giant (HD~222093) lies two orders of magnitudes brighter than typical, and three stars lie offscale on the low end. A likely explanation for the difference in the morphology of our figure versus Martínez et al.'s is our improved treatment of interstellar extinction. If we artificially set our extinctions to zero, the figure's morphology qualitatively matches that of Martinez et al.'s. Despite our lower spectral resolution compared to IUE's, continuum subtraction is too minor to contribute significant error.

Another giant, HD~126327, lies more than an order of magnitude lower than the line but also was caught flaring in H$\alpha$ (Fig. \ref{fig:H_Mg_mags}). This might indicate that stormy events in the photosphere and lower chromosphere might temporarily disrupt the middle chromosphere where Mg2800 arises. Fig.~\ref{fig:low_high_basal} elucidates the fact that stars lying close to the green line in Fig.~\ref{fig:basal_Mg} in fact have higher Mg II $\lambda$2800 flux compared to stars lying way below the same green line in Fig.~\ref{fig:basal_Mg}. 

\begin{figure}
\centering
\includegraphics[width=8.6cm]{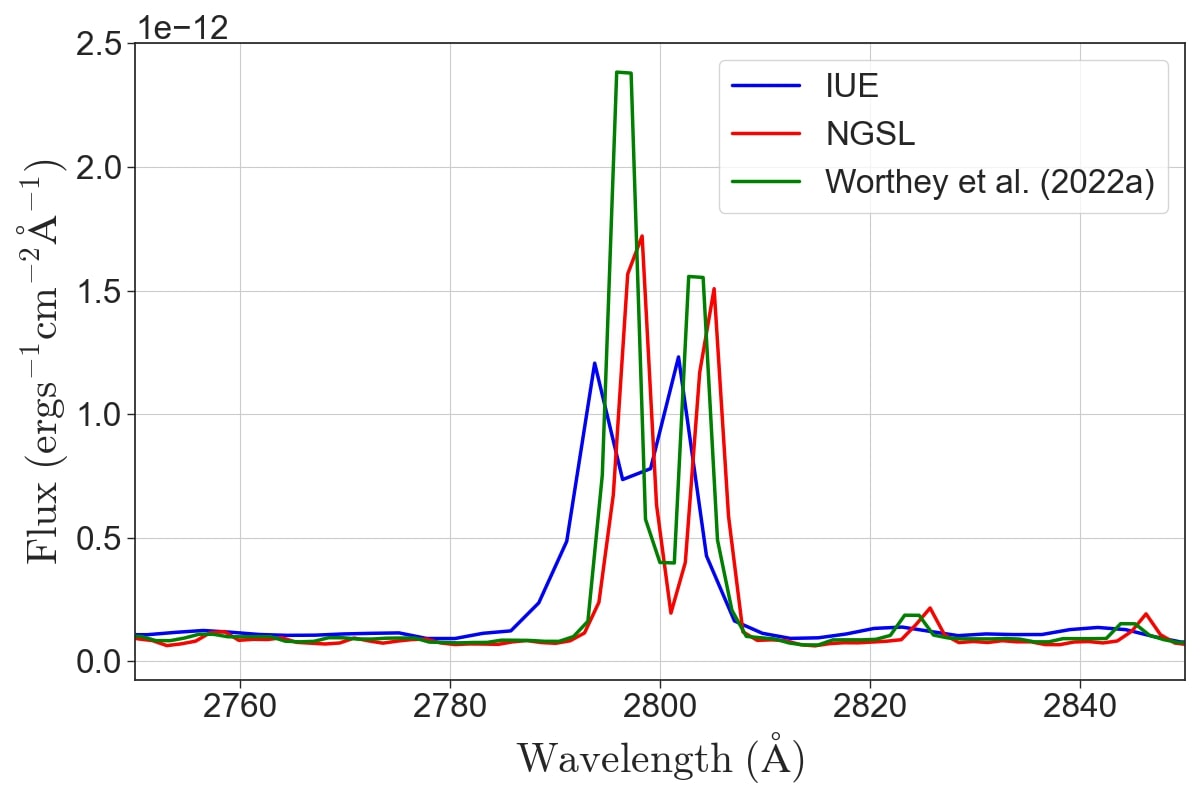}
\caption{Mg II 2800 feature in HD~102212 as observed by IUE (blue), in the NGSL (red), and by \cite{2022stis.rept....5W} (green). The IUE spectrum is at lower resolution compared to \cite{2022stis.rept....5W} and NGSL. }
\label{fig:mg_var}
\end{figure}

Fig.~\ref{fig:mg_var} shows variation in the MgII 2800 spectral lines using observations from International Ultraviolet Explorer (IUE), NGSL, and \cite{2022stis.rept....5W} for the single star HD~102212. The observations were made in 1997, 2002, and 2021 for IUE, NGSL, and \cite{2022stis.rept....5W} respectively. Mg2800 values for the three cases are -1.49$\pm$0.05, -1.81$\pm$0.003, and -2.26$\pm$0.008 for IUE, NGSL, and \cite{2022stis.rept....5W} respectively. The errors in Mg2800 values are calculated by taking into consideration the errors in flux at each pixel value and then propagating these errors while calculating Mg2800 values. Even admitting a few percent additional fluxing error, it is statistically certain that Mg2800 values show a temporal variation in HD~102212. 

Add this to HD~232078, a similar long-period variable listed in $\S$\ref{subsec:note_obj} that is probably also variable in Mg2800.

The sun is known to have a $\sim$7\% Mg2800 variation that correlates with the magnetic activity cycle \citep{1993JGR....9812809D}. \cite{2008A&A...483..903B} report cyclic chromospheric activity in HD~22049 and HD~128621 using IUE spectral data. At visible wavelengths, some studies show overall variation in chromospheric activity from CaII H \& K lines. \cite{1998ASPC..154..153B} report that 85\% of stars in the 40-year HK Project at Mount Wilson Observatory showed either periodic (60\%) or aperiodic (25\%) variation in chromospheric activity. Temporal variation possibly separates magnetically-driven chromospheric heating, which can be expected to be cyclic, from acoustic wave-driven heating, which might be expected to be steadier. In this regard, HD~102212 is not an apt test case because it is a long-period variable star likely to experience considerable ``activity variability'' in its gaseous envelope.

\section{Acknowledgements}
We acknowledge with thanks the variable star observations from the AAVSO International Database contributed by observers worldwide and used in this research. This work is based on observations made with the NASA/ESA Hubble Space Telescope, program GO 16188, \url{https://dx.doi.org/10.17909/t9-d42d-z465}. Support for this work was provided by NASA through grant number HST-GO-16188.001-A from the Space Telescope Science Institute. STScI is operated by the Association of Universities for Research in Astronomy, Inc. under NASA contract NAS 5-26555. This research has made use of the SIMBAD database, operated at CDS, Strasbourg, France.

\clearpage
\bibliography{HLowLib}{}
\bibliographystyle{aasjournal}

\end{document}